\documentclass[journal]{IEEEtran}
\ifCLASSINFOpdf
\else
\fi

\usepackage{ifpdf}
\usepackage[pdftex]{graphicx}
\usepackage[cmex10]{amsmath}
\usepackage{subfig}
\usepackage{amssymb}
\usepackage{mathrsfs}
\usepackage{amsmath}
\usepackage{indentfirst}
\usepackage{booktabs}
\usepackage{multirow}
\usepackage{graphicx}
\usepackage{epstopdf}
\usepackage{fixltx2e}
\usepackage{soul}
\usepackage{units}
\usepackage{mathtools}

\usepackage{caption}
\usepackage{epsfig,graphicx,amssymb,amsmath}
\usepackage{subfig}
\usepackage{diagbox}
\usepackage{color}
\usepackage{setspace}

\usepackage{amsthm}
\usepackage{algorithm}
\usepackage[noend]{algpseudocode}
\usepackage{booktabs}
\usepackage{diagbox}
\usepackage{float}
\usepackage{epstopdf}
\usepackage{ragged2e}
\usepackage{stfloats}
\renewcommand{\raggedright}{\leftskip=0pt \rightskip=0pt plus 0cm}
\usepackage{amsmath}
\usepackage{amssymb}
\usepackage{amsthm}
\usepackage[labelformat=simple]{subcaption}

\ifCLASSINFOpdf
\else

\fi

\usepackage{xcolor}

\usepackage{xpatch}

\usepackage[colorlinks,
            linkcolor=blue,
            anchorcolor=blue,
            citecolor=blue]{hyperref}
\usepackage[noabbrev]{cleveref}

\begin{document}
%
\title{Smart Interrupted Routing Based on Multi-head Attention Mask Mechanism-Driven MARL in Software-defined UASNs}

\author{Zhenyu Wang,~\IEEEmembership{Student member,~IEEE},
Chuan Lin,~\IEEEmembership{Member,~IEEE},
Guangjie Han,~\IEEEmembership{Fellow,~IEEE},
Shengchao Zhu, ~\IEEEmembership{Student member,~IEEE},
Ruoyuan Wu,
Tongwei Zhang,~\IEEEmembership{Member,~IEEE},
Jialu Tian

\thanks{Received 11 July 2025; revised 1 February 2026; accepted 5 February 2026.
Date of publication 10 February 2026; date of current version 5 June 2026.
This work was supported in part by the Joint Funds of the National
Natural Science Foundation of China under Grant 62572107, Grant U22A2011, and Grant 62572169, in part by the
Liaoning Provincial Doctoral Research Start-up Fund Award under Grant 2025-BS-0100, in part by the Fundamental Research Funds for the Central Universities under Grant B250203011, in part by the Open Fund of State Key Laboratory of Acoustics and Marine Information under Grant SKLA202501 and
in part by the National Training Program of Innovation and Entrepreneurship for
Undergraduates under Grant 202510145028, Grant S202510145235, Grant 260210, and Grant 260209.}
\thanks{\emph{Corresponding authors: Chuan Lin \& Guangjie Han}} 
\thanks{Zhenyu Wang, Chuan Lin, Ruoyuan Wu, Jialu Tian are with Software College, Northeastern University, Shenyang, China. (e-mails: larrywang1019@outlook.com; chuanlin1988@gmail.com;  1462938233@qq.com; 15942623988@163.com).}
\thanks{Guangjie Han is with Key Laboratory of Maritime Intelligent Network Information Technology, Ministry of Education, Hohai University. (e-mail: hanguangjie@gmail.com).}
\thanks{Shengchao Zhu is with State Key Laboratory of Acoustics and Marine Information, Chinese Academy of Sciences, Beijing 100190, China. (e-mail: zhushengchao77@gmail.com).}
\thanks{Tongwei Zhang is with the Institute of Marine Science and Technology, Shandong University, Qingdao 266237, China. (e-mail: twzhang@sdu.edu.cn).}
\thanks{Digital Object Identifier 10.1109/TMC.2026.3663303.}

 	
}

\markboth{Journal of \LaTeX\ Class Files,~Vol.~XX, No.~X, XXX~XXXX}%
{Shell \MakeLowercase{\textit{et al.}}: Bare Demo of IEEEtran.cls for IEEE Journals}
%

\IEEEtitleabstractindextext{%
	\begin{abstract}

Routing-driven timely data collection in Underwater Acoustic Sensor Networks (UASNs) is crucial for marine environmental monitoring, disaster warning, and underwater resource exploration, etc.
However, harsh underwater conditions, including high delays, limited bandwidth, and dynamic topologies, make efficient routing decisions challenging in UASNs.
In this paper, we propose a smart interrupted routing scheme for UASNs to address dynamic underwater challenges.
We first model underwater noise influences from real underwater routing features, e.g., turbulence and storms.
We then propose a Software-Defined Networking (SDN)-based Interrupted Software-defined UASNs Reinforcement Learning (ISURL) framework, which ensures adaptive routing through dynamical failure handling (e.g., energy depletion of sensor nodes or link instability) and real-time interrupted recovery.
Based on ISURL, we propose the MA-MAPPO algorithm, integrating multi-head attention mask mechanism with MAPPO to filter out infeasible actions and streamline training.
Furthermore, to support interrupted data routing in UASNs, we introduce MA-MAPPO\_i, MA-MAPPO with interrupted policy, to enable smart interrupted routing decisions in UASNs. 
The evaluations demonstrate that our proposed routing scheme achieves exact underwater data routing decisions with faster convergence speed and lower routing delays than existing approaches.

\end{abstract}
	
	}

\maketitle
\IEEEdisplaynontitleabstractindextext
\IEEEpeerreviewmaketitle

\IEEEpeerreviewmaketitle

\section{Introduction}

\label{sec:introduction}

\IEEEPARstart {M}ARINE resources play a vital role in human development, making their protection and sustainable utilization critical global priorities. 
Underwater Acoustic Sensor Networks (UASNs) have emerged as essential technologies to monitor and manage these resources \cite{WOS:001314081200001}.
Normally, the data are collected at the underwater sensors and are delivered to the cluster head or sink nodes based on active data routing protocols.
However, apart from the unique communication features of UASNs in the underwater environment, such as large propagation delay \cite{WOS:001005658000003}, limited bandwidth \cite{WOS:001353555600001}, the following two limitations have to be confronted and present significant challenges to efficient data routing in UASNs:

\textbf{Complexity of adapting to dynamic environments:} variations in acoustic channel conditions, node mobility due to water currents, and the energy constraints of underwater sensors exacerbate routing instability.
These dynamic factors cause unpredictable changes in the network topology \cite{WOS:001457783700017}, making it difficult for conventional routing algorithms to maintain stable and reliable communication paths \cite{WOS:001317694500023}. 
The unpredictable and continuously changing conditions of the underwater environment demand routing strategies capable of intelligent adaptation in real-time, a capability that traditional methods often fail to provide.

\textbf{Lack of real-time updates to network strategies: }conventional approaches often rely on static or heuristic-based routing decisions, which fail to optimally utilize network resources in real-time. 
For instance, many existing schemes are based on predefined routing metrics, such as hop count or link quality, that do not adapt quickly to sudden changes in the environment.
In large-scale UASNs, such as those deployed for ocean-wide environmental monitoring, nodes make routing decisions based on their local knowledge without considering global network states like congestion or optimal data flow paths. This leads to routing inefficiencies, where data may traverse suboptimal paths, increasing latency and reducing overall throughput.

Totally, traditional routing policies designed for terrestrial wireless networks struggle to adapt to harsh underwater conditions, which require more robust and adaptive solutions tailored to UASNs \cite{WOS:001280054300001}.
Recent studies have explored various aspects of Multi-agent Reinforcement Learning (MARL) and Software-defined Networking (SDN) in data routing in UASNs \cite{WOS:001459530400009}\cite{WOS:001470589200001}.
MARL-based routing approaches leverage deep reinforcement learning techniques to optimize routing selection by continuously learning from the underwater network environment \cite{WOS:001166992300025}.
These approaches have demonstrated improved adaptability and robustness against underwater channel variations. 
On the other hand, SDN-based UASNs frameworks have introduced centralized control mechanisms that enhance network flexibility and scalability.
By decoupling the control and data planes, SDN enables dynamic route optimization and efficient resource allocation. 
Further, in some of recent research products in \cite{WOS:001340385300013}\cite{WOS:001361959500001}, some hybrid solutions have also been proposed that integrate MARL and SDN to leverage their complementary advantages. 
By incorporating MARL, the system enables multiple agents, sensor nodes, to collaboratively learn and adapt to complex underwater environments, further enhancing the network's robustness and scalability \cite{WOS:001142681800640}.
These studies have shown promising results in improving network performance, reducing end-to-end latency, and optimizing energy consumption.

However, despite advances in routing strategies based on MARL and SDN, the following critical issues remain unresolved when integrating these techniques into UASNs; 
1) Traditional MARL algorithms often suffer from slow convergence in highly dynamic environments, making real-time adaptation challenging \cite{WOS:001283796700001};
2) In addition, traditional MARLs consume additional communication overhead for online training, which is intolerable in communication resource-constrained UASNs;
3) Meanwhile, SDN-based network management solutions have to face challenges, including control overhead and centralized failure risks, which can undermine network robustness;
4) Furthermore, the integration of MARL and SDN in UASNs raises new concerns, such as ensuring efficient coordination between distributed agents and centralized controllers, optimizing trade-offs between learning efficiency and network stability, and addressing security vulnerabilities.

This study proposes an SDN-based MARL framework for UASNs in dynamic underwater environments.
Unlike traditional SDN-enabled UASNs, our proposed framework features a three-tier functional stratification: the global network deployment layer, the local view synchronization layer, and the routing implementation layer.
This hierarchical framework tackles traditional MARL limitations (slow convergence, high resource use) through adaptive real-time routing optimization.
The global deployment layer partitions the network into subnets to reduce state space, addressing MARL's slow convergence.
The local view synchronization layer and the routing implementation layer employ Centralized Training Decentralized Execution (CTDE) based MARL for regional training with limited view updates to minimize communication overhead.

Building on this framework, we develop a MARL-driven routing algorithm using multi-head attention mask mechanism for dynamic adaptation to underwater channel variations, reducing centralized training needs.
Note that, inspired by recent routing policies in UASNs, we integrate the interrupted policy in the proposed MARL-driven routing algorithm, which allows real-time recovery from routing failures caused by energy depletion or link instability, thus addressing challenges related to control overhead and centralized failures in SDN-based networks.

Our approach novelly deep-integrates the SDN framework with MARL algorithms, moving beyond basic technique combinations.
Unlike conventional SDN-enabled UASNs relying on periodic controller updates, our framework enables collaborative global-local decision-making. 
The proposed algorithm in this work fundamentally improves the traditional MARL algorithm by incorporating a multi-head attention mask to preemptively filter infeasible actions and an interrupted policy for dynamic failure recovery.
This tight architectural-algorithmic coupling ensures superior adaptability and efficiency in highly dynamic underwater environments.

In total, the main contributions of this work can be summarized by the following:

\begin{enumerate}[]
	\item  We utilize the SDN technique to redefine the framework of UASNs and propose ISURL, an interrupted software-defined UASNs reinforcement learning framework supporting scalable MARL-driven routing decision;
	\item For ISURL, we propose a Multi-head Attention Mask Mechanism-based MAPPO (MA-MAPPO) algorithm, to support fine-grained routing training and decision;
     \item We introduce a smart interrupted policy to MA-MAPPO, presenting MA-MAPPO\_i to support intelligent underwater routing decisions for UASNs under complex underwater conditions. 
\end{enumerate}

The rest of this paper is organized as follows: Related works about this work are surveyed in Sec. \ref{Section:2}; 
Sec. \ref{Section:3} introduces the system model; 
Sec. \ref{Section:4} displays the proposed ISURL, while Sec. \ref{Section:5} presents the proposed interrupted routing decision scheme;
The evaluation results are showcased in Sec. \ref{Section:6}; 
Sec. \ref{Section:7} concludes the paper and discusses the future research directions derived from this paper.

\section{Related Works}\label{Section:2}
In this section, we showcase recent research progress in related research areas, which are specifically grouped into the following three categories: 1) software-defined UASNs, 2) routing in UASNs, and 3) routing based on RL / MARL.

\subsection{Software-defined UASNs}\label{Section:2-1}
In Software-defined Underwater Acoustic Sensor Networks (UASNs), routing separates control and data planes, enabling centralized control and dynamic optimization.
This framework is particularly useful in dynamic underwater environments, supporting real-time adjustments and efficient protocol management for multi-agent systems.

In \cite{WOS:001470589200001}, a Software-Defined Underwater Acoustic Sensor Network (SD-UASN) architecture is proposed.
SDN optimizes data collection by decoupling the control and data planes, enabling centralized control for better network monitoring.
Key innovations include dynamic node selection, adaptive topology management, and energy-efficient data aggregation. 
These enhancements improve network efficiency, reduce overhead, and ensure reliable communication in large-scale underwater monitoring applications, where real-time adaptability is essential.

In \cite{WOS:000976244700034}, authors present an ``SDN+AI"-based framework for software-defined Underwater Internet of Things (IoUT) to tackle load balancing and QoS issues. 
By integrating SDN to separate data and control planes, the framework boosts network scalability and flexibility.
A switch-migration-based multi-controller load-balancing strategy, CASM, is proposed to optimize performance by reducing response time and control path latency. 
Leveraging SDN controllers' global view, the QoS-aware SQAR protocol using reinforcement learning enables intelligent route selection for various IoUT services. CASM achieves effective load balancing, and SQAR outperforms existing protocols in QoS satisfaction, energy efficiency, and convergence. The framework sustains a QoS violation rate under 5\% and a load - balancing rate over 90\% in real-time.

\subsection{Routing in UASNs}\label{Section:2-2}

UASNs struggle with bandwidth limitations, signal delays, and node movement from currents. 
Advanced routing combines acoustic modeling and real-time topology analysis to optimize paths. 
These systems continuously monitor connections, adjusting routes to avoid disruptions, ensuring reliable data transmission in challenging underwater conditions.

In \cite{ WOS:001470473300001}, authors propose an approach utilizing phantom nodes and secure zones to improve the privacy and reliability of source node locations and prevent location leakage, and mitigate passive attacks. 
The Q-learning algorithm has been employed for dynamic selection of relay nodes, optimizing energy consumption and reducing delays. 
Additionally, non-uniform clustering strategies have been introduced for deploying auxiliary nodes that transmit interference signals, effectively protecting against active threats while ensuring secure and efficient routing in UASNs.

In \cite{WOS:001445070500033}, authors propose a topology evolution model for UASNs that leverages scale-free networks to improve robustness and energy efficiency. 
The model integrates factors like flow load, energy consumption, and distance into the preferential attachment mechanism to balance network load and enhance attack resistance. 
Additionally, a network flow adjustment algorithm is introduced, which jointly selects paths and power levels based on residual energy, ensuring energy consumption is balanced across nodes. 
This approach extends network lifetime by optimizing routing and energy management, providing a more efficient solution for UASN routing under constrained conditions.

\subsection{Routing based on RL / MARL} \label{Section:2-3}

In network routing, RL and MARL tackle path-planning complexities in dynamic networks.
RL agents optimize paths through reward-based learning, adapting to real-time flow, link quality, and congestion to maximize throughput or minimize delay. 
MARL extends this with decentralized agent collaboration/competition, sharing link availability, and traffic demands to balance loads and dynamically reroute packets.
Both integrate real-time learning with distributed designs, excelling in dynamic path planning for modern networks that need continuous optimization and resilience against topology changes or traffic surges.

In \cite{WOS:001459530400009}, authors propose STGR to address routing challenges in UUV networks.
STGR is a routing protocol integrating a Spatial-Temporal Graph model with Q-learning. STGR uses a distributed STG model to represent evolving neighbor relationships, link quality, and connectivity duration, and employs a Q-learning-based forwarder selection algorithm for dynamic adaptation to the underwater environment.

In \cite{WOS:001483850200017}, authors propose frameworks that divide road networks into sub-regions, enabling more efficient routing. 
Additionally, techniques such as low-dimensional global states, trajectory collection, and cooperative actor networks have been employed to improve multi-agent cooperation. 
These innovations allow for effective asynchronous route planning, with recent evaluations showing that such models outperform traditional planning methods on both synthetic and real-world road networks.

In \cite{WOS:001166992300025}, authors have focused on enhancing IoUT network reliability by addressing node failures and communication issues. 
Multi-agent systems, utilizing reinforcement learning, allow autonomous underwater vehicles (AUV) to adapt to dynamic conditions and efficiently repair faulty nodes. 
Coordination between AUVs optimizes coverage and repair efficiency, while area information entropy reduces redundancy, improving overall node repair performance and ensuring stable IoUT operations.

Underwater routing in UASNs encounters significant challenges in maintaining adaptability to dynamic marine conditions while ensuring reliable data transmission and system resilience. 
To overcome these limitations, we present an innovative SDN-enabled MARL framework designed to enhance network scalability and operational efficiency. 
This approach not only improves environmental adaptability for diverse mission requirements but also incorporates an advanced routing mechanism to significantly strengthen the system's overall scalability and performance in underwater environments.

\section{System Model}\label{Section:3}
In this section, to accurately simulate data routing in UASNs with real underwater environments, we model UASNs with marine noise.
Meanwhile, we demonstrate how to transform the adaptive data routing decision process into a Markov Decision Process (MDP) to optimize data transmission efficiency.

\subsection{Underwater Noise Modeling in UASNs}\label{Section:3-1}

The underwater noise originates from four parts: underwater vehicle noise, thermal noise, turbulence noise, and storm noise.

\textbf{Underwater vehicle noise} mainly stems from the vibration of the vehicles (e.g, ships, submarines, and AUVs\cite{WOS:001232438400001}) and the impact of water flow.
The noise level \( SL_{vehicle}\) is influenced by the vehicle's speed and the density of vehicles (the number of vehicles per unit volume), and can be calculated using Eq. \ref{EQ1}:

\begin{equation}
\label{EQ1}
    SL_{vehicle} = 186 - 20\lg f + 6\lg\left(\frac{v_h}{6.18}\right) + 10\lg \rho_N,
\end{equation}
where \( f \) represents the frequency of the vehicle's sonic wave, \( v_h \) represents the speed of the vehicle, and \( \rho_N \) represents the vehicle density.

\textbf{Thermal noise} originates from the thermal motion of electrons (e.g., resistors and feed lines\cite{WOS:001126316600024}) in communication equipment components. The noise power spectral density \( S_n(f) \) is given by Eq. \ref{EQ2}:

\begin{equation}
\label{EQ2}
    S_n(f) = 4k_B T R,
\end{equation}
where \( k_B \) represents the Boltzmann constant (\( 1.38 \times 10^{-23} \) J/K), \( T \) represents the absolute temperature and \( R \) is the resistance value. The total noise power within a specific bandwidth \( B \) can be calculated by Eq. \ref{EQ3}.

\begin{equation}
\label{EQ3}
    P_{thermal} = S_n(f) \cdot  B.
\end{equation}

Particularly, the thermal noise level \( SL_{thermal} \) in decibels (dB) can then be calculated using a reference power \( P_0 \) (typically \( 1 \text{pW} \) in underwater acoustics):

\begin{equation}
\label{EQ4}
    SL_{thermal} = 10 \lg \left( \frac{P_{thermal}}{P_0} \right).
\end{equation}

\textbf{Turbulence noise} is mainly caused by ocean turbulence affecting sensor nodes and their external components.
The turbulence noise level \( SL_{turbulence} \) is given by Eq. \ref{EQ5}:

\begin{equation}
\label{EQ5}
    SL_{turbulence} = SL_{base} + 20\lg U_{turb}(t),
\end{equation}
where \( SL_{base} \) is the base noise level 
and \( U_{turb}(t) \) represents the ocean turbulence speed at time \(t\).

\textbf{Storm noise} is related to wind speed and sea conditions \cite{WOS:001324212800001}. The storm noise level $SL_{storm}$ can be represented by Eq. \ref{EQ6}:

\begin{equation}
\label{EQ6}
\begin{split}
     SL_{storm} = 55 - 6\lg & \Bigg[\Bigg(\frac{f}{400}\Bigg)^2  + 1\Bigg] \\ & + \Bigg(18 + \frac{U_{wind}(t)}{4}\Bigg)\lg\Bigg(\frac{U_{wind}(t)}{10}\Bigg),
\end{split}
\end{equation}
where \( f \) represents frequency, \( U_{wind}(t) \) represents the wind speed at the time \(t\). 


Given that the factors of turbulence and storm, such as speed and the amplitude of fluctuations, are changing in real time, the simulation of \( U_{turb}(t)\) and \(U_{wind}(t)\) incorporates real-time oceanic data from NOAA \cite{WOS:001185429800001}.

To obtain the total underwater noise level, we integrate the noise levels from all sources. 
Assuming that the noise sources are independent, the total noise level \(SPL_{total}\) can be calculated by Eq. \ref{EQ7}.

\begin{equation}
\label{EQ7}
SPL_{total} = 10 \lg \left( \sum_{i=1}^{n} 10^{\frac{SL_i}{10}} \right),
\end{equation}
where \(SL_i\) is the noise level of the \(i\)-th noise source, and \(n\) is the total number of noise sources.

\subsection{Markov Decision Process Modeling for Routing in UASNs}\label{Section:3-2}

In the routing decision-making process in UASNs, the environment can be defined as a MDP, as shown in Eq. \ref{EQ8}. The MDP consists of five main components: the state space \(\mathcal{S}\), the action space \(\mathcal{A}\), the state transition probabilities \(\mathcal{P}\), the reward function \(\mathcal{R}\), and the discount factor \(\gamma\):

\begin{equation}
\label{EQ8}
\mathcal{M} = (\mathcal{S}, \mathcal{A}, \mathcal{P}, \mathcal{R}, \gamma).
\end{equation}

\textbf{State space \(\mathcal{S}\)} includes node positions, view of adjacent nodes, noise influence, and communication quality. 
In this work, \(\mathcal{S}\) can be represented as \(\mathcal{S} = \{s_1, \ldots, s_N\}\), where \(s_i = (p_i, v_i, SPL_i, c_i)\). 
In particular, \(N\) denotes the number of nodes, \(p_i\) represents the position coordinates of node \(i\), \(v_i\) represents the adjacency view of node \(i\), \({SPL}_i\) signifies noise influence experienced by node \(i\), and \(c_i\) represents current communication quality of links associated with node \(i\).

\textbf{Action space \(\mathcal{A}\)} is the set of all possible actions for nodes, denoted as \(\mathcal{A} = \{a_1, \ldots, a_N\}\). 
Each action \(a_i\) is an action vector determining a sensor node's next action, like choosing a subsequent node or remaining stationary. 

\textbf{State transition probability \(\mathcal{P}\)} models changes in states influenced by node positions, actions, and external noise. It is represented as \( P(s'|s,a) = Pr(s' = f(s,a)) \), with \( f(s,a) \) being the state transition function.

\textbf{Discount factor} \(\gamma \in [0,1]\), balances the trade-off between immediate rewards and long-term benefits like stable communication, minimal signal attenuation, and energy efficiency in policy evaluation.

\textbf{Reward function \(\mathcal{R}\)} will be detailed in Sec. \ref{Section:5}, where we consider multiple real factors, e.g., packet forwarding, underwater noise (detailed in Sec. \ref{Section:3}), hop count, and propagation delay to ensure high long-term communication efficiency.

\section{Overview of Proposed ISURL}\label{Section:4}

In this section, we present an overview for the proposed \textbf{I}nterrupted \textbf{S}oftware-defined \textbf{U}ASNs \textbf{R}einforcement \textbf{L}earning (ISURL) framework.

\begin{figure}[bth]
	\centering
	\includegraphics[width=0.8\linewidth]{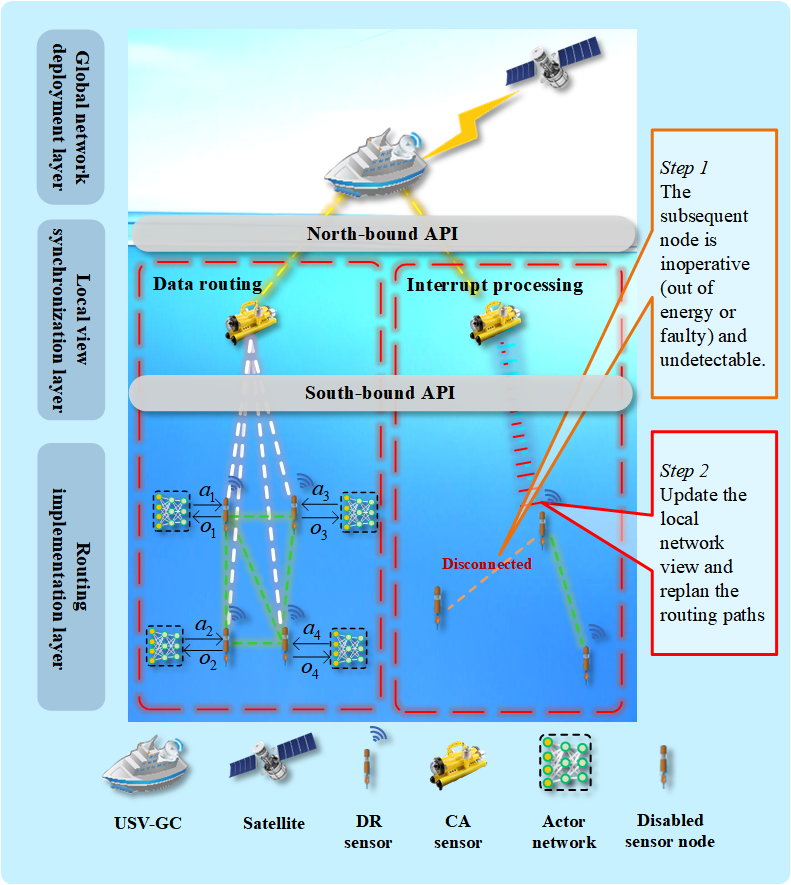}
	\caption{Framework of proposed ISURL}
	\label{fig1}
\end{figure}

As shown in Fig. \ref{fig1}, ISURL is based on the SDN technique to divide UASNs into three functional layers: global network deployment layer for macroscopic network planning and SDN-based control, local view synchronization layer for regional coordination and routing decision-making, and routing execution layer for real-time data forwarding and interruption handling. In particular, we introduce CTDE-based MARL into ISURL, which enables cooperative data routing for sensor nodes in hierarchical UASNs.

\textbf{Global network deployment layer: }
ISURL's global network deployment layer corresponds to the SDN application plane, leveraging the SDN northbound API to optimize task completion while maintaining a global network view.
Within this layer, the Unmanned Surface Vessel-Global Coordinator (USV-GC) employs centralized training in a CTDE-based MARL framework to partition the global UASNs into multiple subnets, each assigned to a subordinate Central Aggregation (CA) sensor, extending SDN's application plane with intelligent decision-making capabilities.
The partitioning of subnets is based on geographic proximity, residual energy, and node density.
Specifically, USV-GC can dynamically adjust subnets upon triggered events like node failure or node offset, and if a subnet’s average energy drops sharply, its size is reduced by reassigning border nodes to adjacent subnets, thus balancing the energy consumption load.
These CA nodes are responsible for data routing and collection within their respective subnets, subsequently assigning specific tasks to the corresponding Data Routing (DR) sensors to implement data collection and routing tasks. 

\textbf{Local view synchronization layer:}
Functioning as an enhanced control plane, this layer utilizes the MARL framework for local training, and the CA sensors transform the global tasks assigned by the global network deployment layer into specific executable tasks for DR sensors in the routing implementation layer. 
These tasks are then dispatched to the subordinate subnets managed by the CA sensors within the routing implementation layer through the SDN southbound API.
Meanwhile, the local network views of the CA sensors are periodically updated via a dedicated SDN southbound API.
CA sensors are not only responsible for managing DR sensors, but also play a crucial role in coordinating between the global network deployment layer and the routing implementation layer. 
Additionally, as the routing endpoint for all DR sensors in the routing implementation layer, the CA sensors receive data packets routed from the DR sensors via the southbound API. 
Consequently, CA sensors typically require high-performance batteries and robust processing and storage capabilities to perform these critical tasks efficiently.

\textbf{Routing implementation layer: }
This layer, composed of DR sensors, constitutes the data plane, where DR sensors collect oceanic data and forward it to CA sensors via routing.
The routing implementation layer is categorized into two functional scenarios: data routing and interrupted processing.

\paragraph{\textbf{Data routing}}
Each DR sensor collects data within its designated area and transmits it to the CA sensor in the local view synchronization layer via routing paths, completing a round of data collection.
The framework supports scalable and logically independent underwater data routing in UASNs, facilitating the deployment of complex routing strategies in the routing implementation layer. 
Note that in Sec. \ref{Section:5} of this paper, a MARL-based routing algorithm is introduced to optimize the routing paths between DR and CA sensors.

\paragraph{\textbf{Interrupt processing}}
In the dynamic marine environment, nodes may fail during routing due to energy depletion or link instability.
When a DR sensor detects that its next-hop node is unreachable, it temporarily buffers the data packet instead of discarding it. 
The DR sensor then sends a request to the CA sensor in the local view synchronization layer, to seek an alternative routing path. 
Upon receiving the request, the CA sensor updates its regional network view, allowing it to assist the DR sensor in recalculating the new route by the same MARL algorithm proposed in Sec. \ref{Section:5}. 
Upon receiving a response, the DR sensor either forwards the buffered data through the newly assigned path or continues buffering if no immediate route is available.

\section{Proposed Interrupted Routing Decision Scheme}\label{Section:5}

\subsection{Proposed MA-MAPPO}\label{Section:5-1}
Among MARL algorithms, the Multi-Agent Proximal Policy Optimization (MAPPO) algorithm has demonstrated superior performance in cooperative multi-agent scenarios.
However, inefficient action exploration and redundant decisions may still hinder convergence speed, which all of MARLs have to confront. 
To address this issue, we propose the Multi-head Attention Mask Mechanism-driven MAPPO (MA-MAPPO), which integrates a multi-head attention mask mechanism into the Actor network of MAPPO for action evaluation.
The key innovation is to filter out infeasible actions through multi-head attention mask mechanism before policy optimization, reducing unnecessary exploration and accelerating convergence.

\begin{figure*}[h!]
    \centering
    \includegraphics[width=0.8\linewidth]{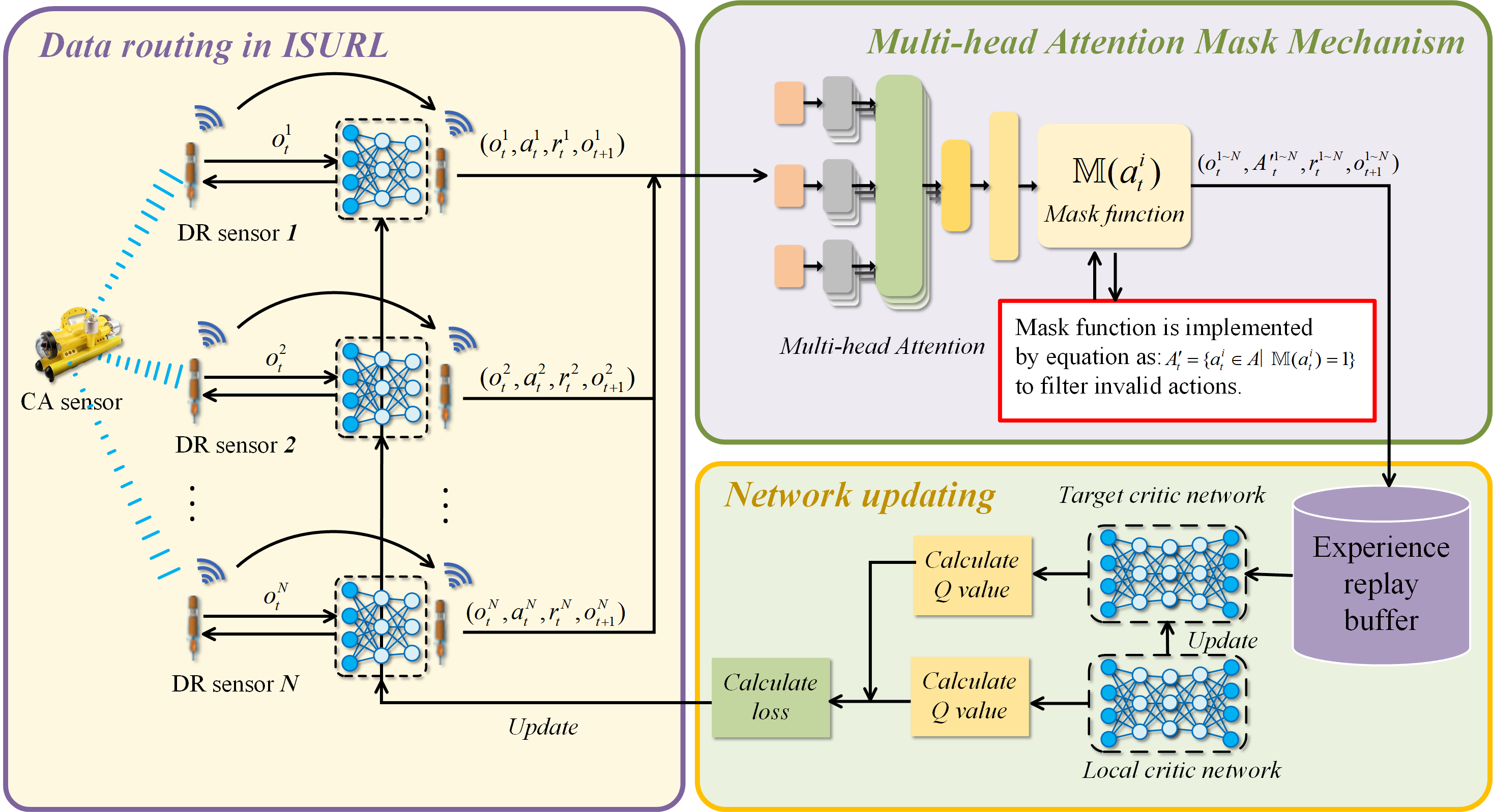}
    \caption{UASNs-based MA-MAPPO framework}
    \label{fig2}
\end{figure*}

As shown in Fig. \ref{fig2}, the core idea of MA-MAPPO is to use a multi-head attention module to estimate the feasibility of each action based on the current state.
Given the state \(s_t^i\) of the agent (node) \(i\) at time \(t\), the attention mechanism computes attention scores as Eq. \ref{EQ9}.

\begin{equation}
\label{EQ9}
    M_t = \text{softmax} \left( \frac{Q_t W_Q (K_t W_K)^T}{\sqrt{d_k}} \right) V_t W_V,
\end{equation}
where \(Q_t = f_Q(s_t)\), \(K_t = f_K(s_t)\), and \(V_t = f_V(s_t)\) are learnable feature representations of the state \(s_t\), \(W_Q, W_K, W_V\) are trainable projection matrices, \(d_k\) is the dimensionality of the key vector.
Attention scores \(M_t\) quantifies the importance of different actions, to construct an adaptive action mask.

To eliminate infeasible or suboptimal actions, we apply a threshold-based action mask. Specifically, we define a threshold \(\tau\) and generate a binary mask \(\mathbb{M}(a_t^i)\):

\begin{equation}
\label{EQ10}
\mathbb{M}(a_t^i) = \begin{cases} 
1, & \text{if } M_t^i \geq \tau \\ 
0, & \text{otherwise}
\end{cases},
\end{equation}
where \(\tau\) is defined as the attention threshold for filtering invalid actions, and its calculation is defined in Eq. \ref{eqtau},
\begin{equation}
\label{eqtau}
    \nabla_{\omega}J(\theta) = \mathbb{E}_{\pi_\theta}\left[\nabla_\omega\tau\cdot\hat{A}^{\pi_{\theta}}(s_t | a_t)\right],
\end{equation}
where \(\nabla_\omega \tau = \tau(1-\tau)\cdot[1,\log(N_S)]\), \(N_S\) is the number of nodes in current network.

The valid action space after applying the mask can be calculated by Eq. \ref{EQ11}:

\begin{equation}
\label{EQ11}
A'_{t} = \{a_t^i \in A \mid \mathbb{M}(a_t^i) = 1 \}.
\end{equation}

By dynamically filtering actions, we ensure that the policy makes high-quality decisions while avoiding unnecessary exploration.

Once the action space is masked, we redefine the policy distribution to account for the reduced set of available actions, i.e., sensor nodes will only forward data packets on valid routing paths in the UASNs routing process.

\begin{equation}
\label{EQ12}
\pi'_{\theta_i}(a_t^i | s_t^i) = \frac{\pi_{\theta_i}(a_t^i | s_t^i) \mathbb{M}(a_t^i)}{\sum_{a' \in A} \pi_{\theta_i}(a' | s_t^i) \mathbb{M}(a')}.
\end{equation}

The policy gradient is then updated using Eq. \ref{EQ13}:

\begin{equation}
\label{EQ13}
    \nabla_{\theta} J(\theta) = \mathbb{E}_{r \sim \pi_{\theta}} \left[ \nabla_{\theta} \log \pi_{\theta} (a_t | s_t) \cdot \hat{A}^{\pi_{\theta}} (s_t | a_t) \right],
\end{equation}
where $\pi_{\theta}$ represents the policy network, $a_t$ represents the action taken at state $s_t$, and $\hat{A}^{\pi_{\theta}}(s_t | a_t)$ represents the estimated advantage function, which is defined as follows:

\begin{equation}
\label{EQ14}
\hat{A}^{\pi_{\theta}}(s_t | a_t) = Q^{\pi_{\theta}}(s_t, a_t) - V^{\pi_{\theta}}(s_t) ,
\end{equation}
where $Q^{\pi_{\theta}}(s_t, a_t)$ represents the action-value function, i.e., the expected return when taking action $a_t$ at state $s_t$ by following policy $\pi_{\theta}$.
In Eq. \ref{EQ14}, $V^{\pi_{\theta}}(s_t)$ represents the state-value function, i.e., the expected return by following $\pi_{\theta}$ at state $s_t$.

In the proposed MA-MAPPO algorithm, the value function update equation is given by Eq. \ref{EQ15}:

\begin{equation}
\label{EQ15}
\nabla_{\phi} J(\phi) = \frac{1}{2} \mathbb{E}_{s \sim \rho^{\pi_{\theta}}} \left[ \left( V^{\pi_{\theta}}(s) - \hat{V}^{\pi_{\theta}}(s) \right)^2 \right],
\end{equation}
where $\phi$ represents the trainable parameters of the value network, $\rho^{\pi_{\theta}}$ represents the state distribution under policy $\pi_{\theta}$, and $\hat{V}^{\pi_{\theta}}(s)$ represents the target value function. The target value function is obtained using Eq. \ref{EQ16}:

\begin{equation}
\label{EQ16}
\hat{V}^{\pi_{\theta}}(s) = r + \gamma \sum_{s'} p(s' | s, a) V^{\pi_{\theta}}(s').
\end{equation}

The Critic network is updated via mean squared error loss using Eq. \ref{EQ17}:

\begin{equation}
\label{EQ17}
L(\phi) = \mathbb{E} \left[ \left( Q_{\phi}(s_t, a_t) - R_t \right)^2 \right],
\end{equation}
where \(Q_{\phi}(s_t, a_t)\) is the estimated value function, and \(R_t\) is the observed reward.

\begin{algorithm}[htb]
    \caption{Proposed MA-MAPPO}
    \label{alg: MA-MAPPO}
    \begin{algorithmic}[1]
        \State \textbf{Initialize:} Policy parameters \(\theta\), Critic parameters \(\phi\), attention weights, replay buffer \(\mathcal{D}\), and environment;
        \For{each training iteration}
            \For{each agent \(i\)}
                \State Observe state \(s_t^i\) and compute attention-based action scores \(M_t\) via Eq. \ref{EQ9};
                \State Apply action mask \(\mathbb{M}(a_t^i)\) and select valid action space \(A_t'\) using Eq. \ref{EQ10} and Eq. \ref{EQ11};
                \State Sample action \(a_t^i\) from masked policy \(\pi'_{\theta_i}(a_t^i | s_t^i)\);
                \State Execute action \(a_t^i\), receive reward \(r_t\), and store \((s_t, a_t, r_t, s_{t+1})\) in replay buffer \(\mathcal{D}\);
            \EndFor
            \State Estimate advantage function \(A_t\) using GAE;
            \State Compute value function loss via Eq. \ref{EQ14} and update \(\phi\);
            \State Compute policy gradient \(\nabla_{\theta} J(\theta)\) via Eq. \ref{EQ13} and perform PPO-Clip update on \(\theta\);
            \State Update the attention threshold \(\tau\) by Eq. \ref{eqtau}.
        \EndFor
    \end{algorithmic}
\end{algorithm}

\begin{figure*}[h!]
  \centering
  \includegraphics[width=0.8\linewidth]{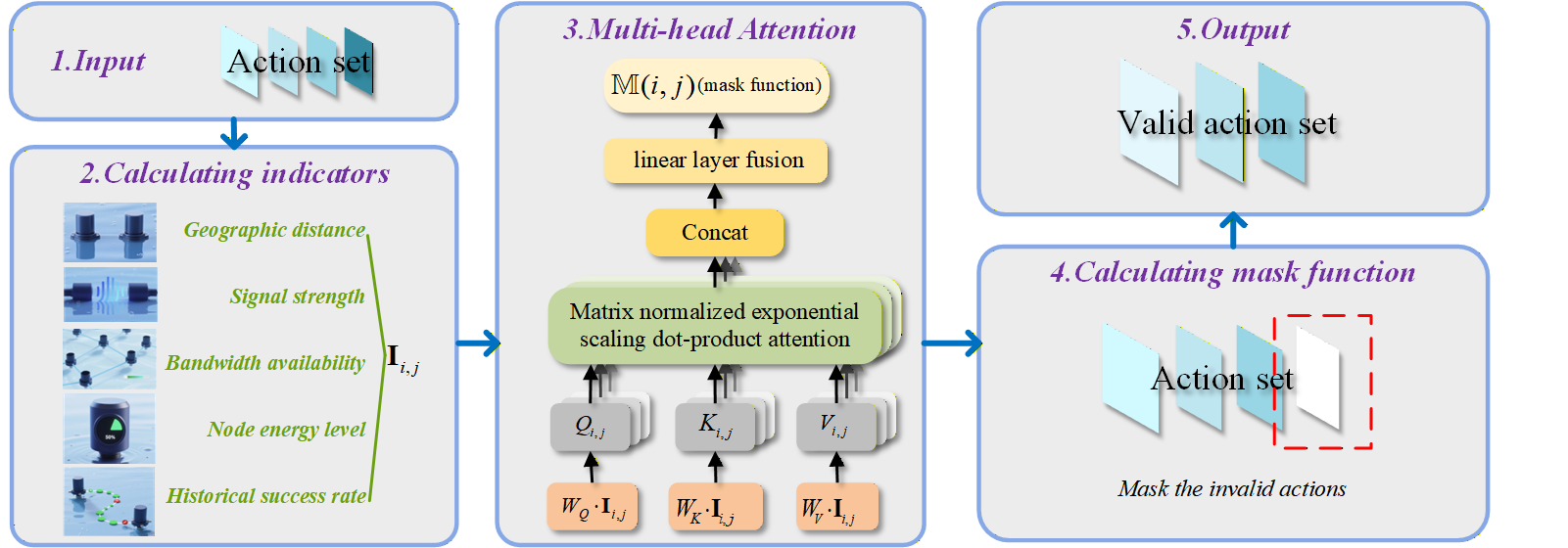}
  \caption{ Mask function implementation process }
  \label{fig3}
\end{figure*}

Totally, we summarize the entire procedure of the proposed MA-MAPPO in Algorithm \ref{alg: MA-MAPPO}. The algorithm starts by initializing the policy parameters, the critic network, and the attention weights (Line 1).
It then iterates through multiple training cycles, each  beginning with collecting trajectory data (Lines 2-7).
During this process, each agent observes its state and computes attention-based action scores (Line 3).
The action mask is applied (Line 4), and a valid action space is selected (Line 5). 
The agent then samples an action from the masked policy (Line 6), executes it, and stores the experience in the replay buffer (Line 7).
Once trajectories are collected, the algorithm estimates the advantage function using the Generalized Advantage Estimation (GAE) method \cite{WOS:001317694500113} (Line 8). 
The Critic network is then updated using the value function loss (Line 9), while the Actor network is optimized using the PPO-Clip objective (Line 10).
The attention threshold for filtering invalid actions is also updated each iteration through Eq. \ref{eqtau} (Line 11).
The process repeats multiple iterations to refine policy performance.

\subsection{Proposed MA-MAPPO\_i for interrupted routing}\label{Section:5-2}

\subsubsection{Network view updating based on MA-MAPPO}\label{Section:5-1-2-1}
\hfill

In MARL based routing decision, node reachability is a crucial factor in routing decision making within dynamic network environments. 
This subsection defines a mask function based on multi-head attention mask mechanism and further integrates it into the MA-MAPPO framework to evaluate the reachability between nodes, thus the network view can be further updated as shown in Fig. \ref{fig3}.

By Eq. \ref{EQ10}, for nodes $i$ and $j$, the corresponding mask function is defined as Eq. \ref{EQ18}.

\begin{equation}
    \label{EQ18}
    \mathbb{M}(i, j).
\end{equation}

Based on Eq. \ref{EQ18}, we introduce additional assessment criteria to refine the accessibility assessment of the nodes in the floating network. 
To ensure a more comprehensive evaluation of connectivity and communication reliability for UASNs, the refined function integrates the following indicators.
We have normalized or utilized other calculation methods to ensure that the obtained indicators are within the range of \([0,1]\), the sum of indicators \(I_{i,j}\) between node \(i\) to node \(j\) can be calculated by the following Eq. \ref{eqgeo} to \ref{eqhis}:

\noindent\textbf{Indicator 1: Geographic distance}

A shorter physical distance between nodes indicates higher accessibility:
\begin{equation}
\label{eqgeo}
I_{i,j} += 1 - \frac{D(i, j)}{D_{\max}},
\end{equation}
where \(D(i, j)\) represents the physical distance between nodes \(i\) and \(j\), and \(D_{\max}\) is the maximum possible distance in the network.

\noindent\textbf{Indicator 2: Signal strength}

A stronger signal improves communication quality and accessibility:
\begin{equation}
I_{i,j} += \frac{S(i, j) - S_{\min}}{S_{\max} - S_{\min}},
\end{equation}
where \(S(i, j)\) represents the signal strength between nodes \(i\) and \(j\), \(S_{\min}\) is the minimum observed signal strength, and \(S_{\max}\) is the maximum observed signal strength.

\noindent\textbf{Indicator 3: Bandwidth availability}

Higher bandwidth enhances data transmission and network connectivity:
\begin{equation}
I_{i,j} += \frac{B(i, j) - B_{\min}}{B_{\max} - B_{\min}},
\end{equation}
where \(B(i, j)\) represents the available bandwidth between nodes \(i\) and \(j\), \(B_{\min}\) is the minimum observed bandwidth, and \(B_{\max}\) is the maximum observed bandwidth.

\noindent\textbf{Indicator 4: Node energy level}

Nodes with higher residual energy maintain stable communication:
\begin{equation}
I_{i,j} += \frac{\min(E(i), E(j)) - E_{\min}}{E_{\max} - E_{\min}},
\end{equation}
where \(E(i)\) and \(E(j)\) represent the residual energy levels of nodes \(i\) and \(j\), respectively, \(E_{\min}\) is the minimum observed energy level, and \(E_{\max}\) is the maximum observed energy level.

\noindent\textbf{Indicator 5: Historical success rate}

A higher success rate reflects a stable and reliable communication history:
\begin{equation}
\label{eqhis}
I_{i,j} += P_{success}(i, j),
\end{equation}
where \(P_{success}(i, j)\) represents the historical probability of successful communication between nodes \(i\) and \(j\).

Here, we employ multi-head attention mechanism to integrate these indicators:
\begin{equation}
\label{EQ24}
M_t = \text{Attention}(Q_{i,j}, K_{i,j}, V_{i,j}),
\end{equation}
where
\begin{equation}
    \begin{cases}
    \label{EQ25}
        Q_{i,j} = W_Q \cdot \mathbf{I}_{i,j} \\
        K_{i,j} = W_K \cdot \mathbf{I}_{i,j} \\
        V_{i,j} = W_V \cdot \mathbf{I}_{i,j}
    \end{cases},
\end{equation}
where \(\mathbf{I}_{i,j}\) represents the feature vector containing all assessment indicators.
The trainable weight matrices \(W_Q\), \(W_K\), and \(W_V\) transform these indicators into query, key, and value vectors for the computation of multi-head attention. 

Afterwards, the vectors and states of sensor nodes will be utilized in MA-MAPPO for training. 
MA-MAPPO first employs the calculated $M_t$ and the node mask function, as defined in Eq. \ref{EQ10}, Eq. \ref{EQ11}, and Eq. \ref{EQ18}, to assess whether a given node can reach other adjacent nodes within its local network. 
Subsequently, the reachability of the node is determined using the masking function defined in Eq. \ref{EQ10}. 
The execution process of how the mask function works is shown in Eq. \ref{fig4}, inputting actor space, and calculating indicators for further evaluations in the mask function as Eq. \ref{EQ18}, and ultimately outputs the valid action space.
This process dynamically constructs a local network view that serves as the basis for further routing decisions.

\subsubsection{Proposed MA-MAPPO with interrupted policy} \label{Section:5-1-2-2}
\hfill

Based on the proposed ISURL, we introduce an interrupted policy to MA-MAPPO (MA-MAPPO\_i), and the proposed MA-MAPPO\_i is executed mainly in the routing implementation layer and the local view synchronization layer as mentioned in Sec. \ref{Section:4}.
As shown in Fig. \ref{fig4}, the implementation of the proposed MA-MAPPO\_i involves two steps:

\begin{figure}[t!]
  \centering
  \includegraphics[width=0.8\linewidth]{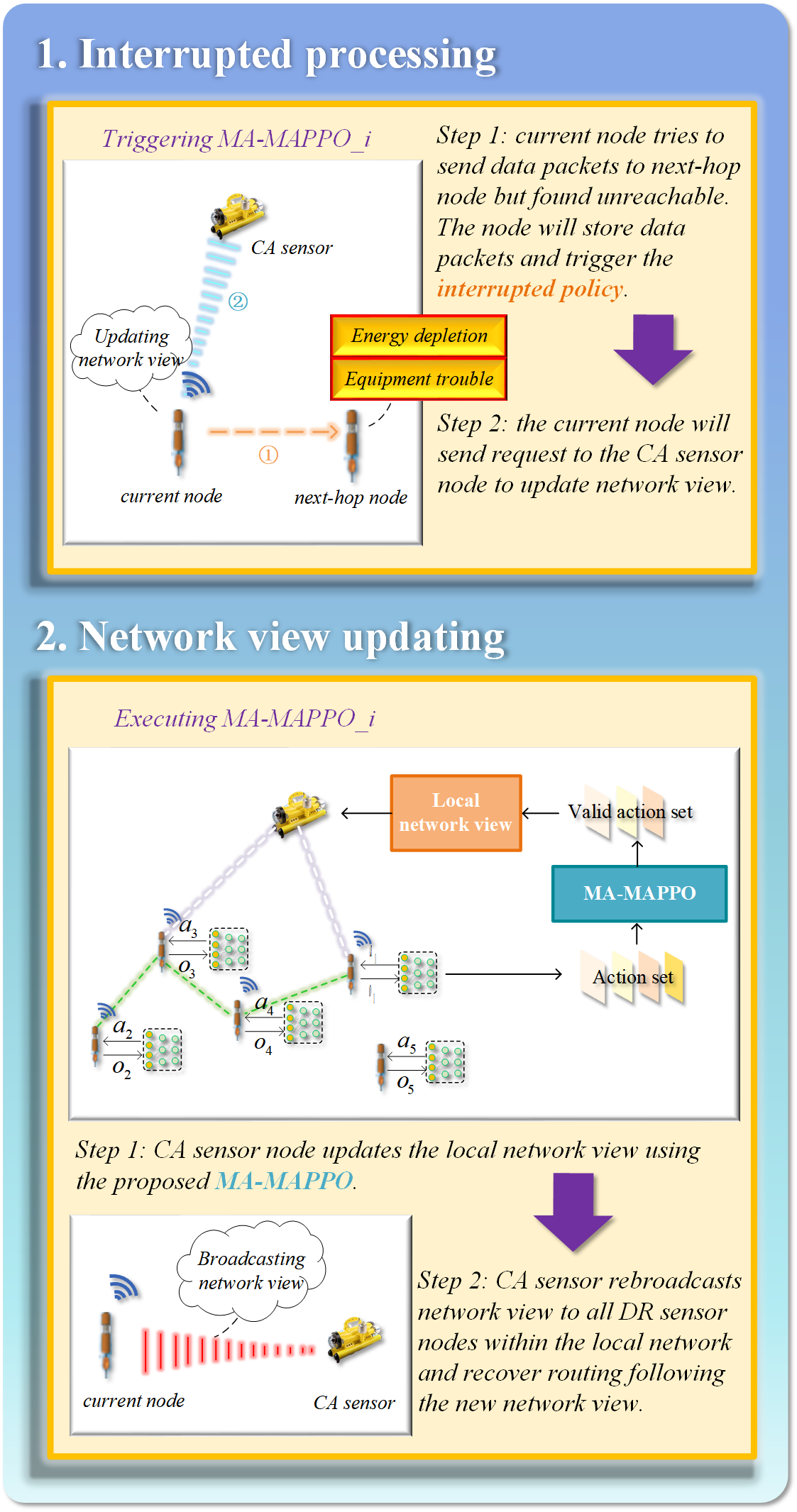}
  \caption{Steps for the proposed MA-MAPPO\_i }
  \label{fig4}
\end{figure}

    \paragraph{Interrupted processing} If the next-hop node of the current node becomes unreachable due to energy depletion or other factors, the current node will immediately interrupt sending data packets to the next-hop node and temporarily store the packets, marking them as ``pending for transmission". 
    The set of cache in each node is detailed in Sec. \ref{Section:6}, an LRU (Least Recently Used) replacement policy is adopted to manage cached packets \cite{WOS:001177637400001}. 
    When the cache reaches its capacity, the oldest packet is discarded to accommodate new incoming data, and the packet loss rate is logged for further analysis.
    Meanwhile, the current node will send a request to the CA sensor node, asking it to dynamically update the local network view using the mask function in MA-MAPPO.
    
    \paragraph{Network view updating} The CA sensor node updates the local network view using the proposed MA-MAPPO-based network view update method and rebroadcasts it to all DR sensor nodes within the local network. 
    Note that, due to the complex and dynamic underwater environment, as well as the relatively large scale of local networks in UASNs, updating the network view requires a significant amount of time and sensor node energy.
    To conserve energy in network view maintenance, the update interval is set to be relatively long, except in cases where an unreachable node causes an ``interruption" in the local network, triggering an immediate update.
    \footnote{
    The path update process avoids single static source-to-sink paths.
    Instead, a dual routing table method for dynamic underwater environments is used, while each node maintains an Outdated routing table and a higher-priority Activated routing table \cite{WOS:000390154400283}.
    During updates, the CA sensor node pushes the Activated routing table to local DR nodes via South-bound API, and the Outdated table is retained briefly within 3 cycles for incomplete old-path packets, then discarded.
    }

The whole process of interrupting the invalid routing process, storing the data packets, updating the network view, and recovering the routing data packets is the implementation of MA-MAPPO\_i, which incorporates the interrupted policy into the proposed MA-MAPPO.
USV-GC periodically aggregates global network states via SDN northbound API for ISURL integration. 
These states are broadcast to CA sensors and incorporated into MA-MAPPO\_i's multi-head attention mechanism, where they are concatenated with local indicators (e.g., geographic distance) in query-key-value calculations (Eq. \ref{EQ24}). 
This enables intelligent action filtering for high-quality routing. During interruptions, if multiple CA sensors report failures, USV-GC coordinates subnet path reassignments to prevent conflicts. 
Such cross-layer synergy ensures MA-MAPPO\_i’s actions align with global optimization goals, enhancing scalability and conflict avoidance in large-scale UASNs.

\subsection{Proposed routing decision based on MA-MAPPO\_i}\label{Section:5-3}

To ensure long-term stable data routing in complex underwater environments, we propose four categories of rewards as shown in Eq. \ref{EQ26}:

\begin{equation}
\label{EQ26}
R_i(s,a) = \vartheta_1 r^{\delta}_i + \vartheta_2 r^{\zeta}_i  + \vartheta_3 r^{\sigma}_i + \vartheta_4 r^{\mu}_i,
\end{equation}
where \(r^{\delta}_i\) is the forwarding reward, \(r^{\zeta}_i\) is the underwater noise reward, \(r^{\sigma}_i\) is the hop count reward, and \(r^{\mu}_i\) is the propagation delay reward. 
And the weights \(\vartheta_1\) to \(\vartheta_4\) in Eq. \ref{EQ26} are optimized through Bayesian optimization during preliminary simulations to balance competing objectives \cite{WOS:001206894305017}.

\textbf{Forwarding reward \(r^{\delta}_i\)} incentivizes nodes to participate in data forwarding and relaying by rewarding successful packet forwarding and penalizing packet loss. The reward function \(r^{\delta}_i\) is defined as Eq. \ref{EQ27}:

\begin{equation}
    \label{EQ27}
    r_i^{\delta} = \begin{cases}
    \gamma_1 \cdot \sqrt{P_{forward}(i)}, & \text{if successful forwarding}, \\
    -\gamma_2 \cdot \sqrt{P_{loss}(i)}, & \text{if packet loss occurs}.
    \end{cases},
\end{equation}
where \(P_{forward}(i)\) and \(P_{loss}(i)\) denote the probabilities of successful forwarding and packet loss for node \(i\), respectively. The weights \(\gamma_1\) and \(\gamma_2\) control the magnitudes of rewards and penalties. Since both \(P_{forward}(i)\) and \(P_{loss}(i)\) are within the interval [0,1], the square root function is applied to amplify their influence.

\textbf{Underwater noise reward \(r^{\zeta}_i\)} is designed to guide data routing to low-noise paths by penalizing nodes in a high-noise environment, thereby improving communication quality. The reward function \(r^{\zeta}_i\) is defined as Eq. \ref{EQ28}:  
\begin{equation}     
    \label{EQ28}     
    r_i^{\zeta} = -\chi \cdot \left( SPL_{total,i} \right)^{\varsigma},
\end{equation}  
where \(SPL_{total, i}\) is the total underwater noise level at node \(i\), calculated by integrating the noise levels from all sources as shown in Eq. \ref{EQ7}, and \(\varsigma\) is a tunable parameter that adjusts the penalization intensity.

\textbf{Hop count reward} \(r^{\sigma}_i\) incentivizes nodes to choose routes with lower hop counts, reducing the probability of packet loss and resource consumption. 
Thus, the hop count reward function \(r^{\sigma}_i\) can be defined as Eq. \ref{EQ29}:  

\begin{equation}  
    \label{EQ29}  
    r_i^{\sigma} = \frac{\alpha}{\beta + H_{hops}(i)}  ,
\end{equation}  
where \(H_{hops}(i)\) represents the number of hops between node \(i\) and the target node, while \(\alpha\) and \(\beta\) are positive scaling parameters that control the reward magnitude and prevent singularities.  
Obviously, \(r^{\sigma}_i\) increases as the hop count decreases, encouraging the system to seek shorter routing paths while allowing flexibility in reward adjustments.  

\textbf{Propagation delay reward} \(r^{\mu}_i\) is designed to motivate the nodes to prioritize routing paths with shorter propagation times, enabling faster data transmission and more efficient utilization of network bandwidth. 
The delay reward function \(r^{\mu}_i\) is defined as Eq. \ref{EQ30}:

\begin{equation}
    \label{EQ30}
    r_i^{\mu} = \omega - \frac{T_{delay}(i) - T_{min}}{T_{max} - T_{min}},
\end{equation}
where \(T_{delay}(i)\) represents the propagation delay between node \(i\) and the target node, \(T_{min}\) and \(T_{max}\) are the minimum and maximum propagation delays in the network, respectively. 
$\omega \in [0,1]$ dynamically adjusts rewards for low-delay paths and penalties for high-delay paths.
This normalized reward function ensures \(r^{\mu}_i\) decreases as \(T_{delay}(i)\) increases, thereby encouraging nodes to select paths with lower delays.

\begin{algorithm}[htb]
    \caption{The proposed interrupted routing scheme based on MA-MAPPO\_i}
    \label{alg: routing}
    \begin{algorithmic}[1]
    \State \textbf{Initialize:} Sensor node set $N$, adjacency matrix $A$, and reward parameters $\theta$
    \While{network is active}
        \For{each sensor node $i$ and $j$ $\in N$ $(i \neq j)$}
            \State Compute $\mathbb{M}(i, j)$ using Eq. \ref{EQ18} and determine connectivity;
            \State Update local network view $V_i$ based on connectivity assessment;
        \EndFor
        \For{each data packet $p_k \in P$}
            \If{ next node $j$ is unreachable}
                \State Store $p_k$ in buffer and mark as pending;
                \State Request CA sensor node to update $V_i$;
                \State Recompute routing path using updated $V_i$;
            \Else
                \State Transmit $p_k$ to next node $j$;
            \EndIf
        \EndFor
        \For{each sensor node $i \in S$}
            \State Compute reward $R_i$ using Eq. \ref{EQ27} - Eq. \ref{EQ30}.
        \EndFor
    \EndWhile
    \end{algorithmic}
\end{algorithm}

The procedure of the proposed interrupted routing scheme is summarized in Algorithm \ref{alg: routing}. 
It initializes sensor node states, the adjacency matrix, and reward parameters (Line 1).
While the network is active, each sensor node iteratively computes its reachability score (Line 3), applies the masking function to determine connectivity (Line 4), and updates its local network view (Line 5).
For each data packet, if the next-hop node is unreachable, the packet is buffered, and the CA node updates the network view and recomputes the route (Lines 8-10). Otherwise, the packet is forwarded (Line 12).
Finally, all routing nodes compute their rewards (Line 14), ensuring adaptive optimization based on real-time network conditions.

\section{Evaluations}\label{Section:6}
To evaluate the availability and efficiency of our proposed routing scheme, taking into account of the proposed SDN-based MARL framework ISURL, we make several comparisons between our proposed MA-MAPPO\_i and some current MARL algorithms in determining the routing path in software-defined UASNs.

\subsection{Simulation setup}\label{Section:6-1}
All the evaluations are conducted on a laptop equipped with an Intel(R) Core(TM) i9-12900H 2.50GHz processor, GeForce RTX 4070 GPU, and 32GB RAM.

We utilize Python 3.9 to program and simulate the underwater evaluation scenarios.
All nodes are set to move randomly due to the unpredictable underwater environment, such as ocean currents. 
The nodes are distributed across the ocean region, with a distance \( d^{\phi}_{{min}} \) of one kilometer between each node, while the source and target nodes are located at least 10 km apart.
In the simulations, three different scenarios (with 64, 125, and 216 nodes, respectively) are taken into account.

\begin{table}[H]
\raggedright
\caption{Simulation parameters}
\label{table1}
\begin{tabular}{lll}
\hline
\textbf{Parameter} & \textbf{Description} & \textbf{Default value} \\ \hline
\multicolumn{3}{l}{\textbf{Model training settings}} \\
\( l_r \) & Learning rate & \( 3e-3 \) \\
\( N_E \) & Number of training rounds & 5000 \\
\( N_h \) & Number of neurons in hidden layer & 64 \\
\( \gamma \) & Discount factor & 0.95 \\
\( \tau \) & Network update coefficient & \( 1e-2 \) \\  \hline
\multicolumn{3}{l}{\textbf{Sensor nodes settings}} \\
\( N_S \) & Number of nodes & [64, 125, 216] \\
\( N_A\) & Number of CA sensors & [1, 2, 4]\\
\( N_D \) & Number of data packets for one task & [500,1000,2000] \\
\( d^{\phi}_{{min}} \) & Minimum distance among DR sensors & 1 km \\
\( d^{\kappa}_{{min}} \) & Minimum distance among CA sensors & 10 km \\
\(N_F\) & Node failure rate & 0.2 \\ \hline
\multicolumn{3}{l}{\textbf{Simulation environment settings}} \\
\( v_s \) & Speed of sound in water & 1531 m/s \\
\( B \) & Available communication bandwidth & 10 kbps\\
\( E_{init} \) & 	
Initial energy level for each node & 1000 J  \\
\( E_{tx} \) & 	
Energy consumption per transmission & 0.5 J  \\
\( E_{rx} \) & 	
Energy consumption for receiving & 0.2 J  \\ 
\( \lambda_{noise} \) & Noise level calculated by Eq. \ref{EQ7} & variable  \\
\( C_{size} \) & 	
Cache size per node & 10 packets  \\
\hline
\end{tabular}
\end{table}

In addition, all evaluations are simulated in a 10 km × 10 km ×
10 km 3D space where the details of the evaluation setting are displayed in Tab. \ref{table1}.
Note that the node failure rate in Tab. \ref{table1} is dynamically modeled by indicators including underwater noises, the speed of waves, the influences of underwater forces, and node remaining energy \cite{CHEN2022108479}.
In each scenario, multiple experiments are repeated, and the results are calculated using samples within a confidence interval of 95\%, ensuring a high level of statistical reliability.
All the parameters in the evaluations are detailed in Tab. \ref{table1}.

Additionally, except for the parameters of the marine environment simulation testing shown in \ref{table1}, the marine databases NOAA \cite{WOS:001185429800001} and global ARGO \cite{roemmich2009} are used in the simulation to achieve a realistic simulation of the marine environment.
Not only do we regard the noise impacts on the system as detailed in \ref{Section:3-1}, to capture the complex interplay of multiple factors in underwater environments, our experimental design but also extends beyond noise modeling to incorporate ocean waves, salinity, and temperature variations of the marine region selected for testing, the latitude range of 3.38°N-16.63°N and the longitude range of 157.37°E-161.41°E.
These elements collectively influence acoustic propagation and node stability.

We first define the noise factor as ${Noise}(t)$, the wave factor as $Wave(t)$, the salinity factor as $Sali(t)$, and the temperature factor as $Temp(t)$ in Eq. \ref{eqs}:
\begin{equation}
\label{eqs}
\begin{cases}
{Noise}(t) = SPL_{total}(t)\\
{Wave}(t) = A_{{wave}} \cdot \sin\left(2\pi f_{{wave}}(t) + \phi\right) \cdot \frac{D(i, j)}{D_{\max}}\\
{Sali}(t),Temp(t) \text{ can be obtained directly}
\end{cases},
\end{equation}
where ${Noise}(t)$ is the noise level calculated by Eq. \ref{EQ7}, $A_{{wave}}$ is the wave amplitude, related to $U_{turb(t)}$ and $U_{wind}(t)$ defined in Eq. \ref{EQ4} and Eq. \ref{EQ5}, $f_{wave}(t)$ is the frequency of the wave, $\phi$ is the phase shift of the wave, \(D(i, j)\) represents the physical distance between nodes \(i\) and \(j\), and \(D_{\max}\) is the maximum possible distance in the network.

All the factors mentioned above are added to the simulation evaluations to make the experimental scene more realistic in reproducing the environment in which the underwater network is located.

\subsection{Results and discussions}\label{Section:6-2}

To measure the proposed MA-MAPPO\_i, we compare MA-MAPPO\_i, with Multi-Agent Proximal Policy Optimization (MAPPO) \cite{WOS:001081924500004}, Multi-Agent Deep Deterministic Policy Gradient (MADDPG) \cite{WOS:001470589200001}, Multi-Agent Twin Delayed Deep Deterministic (MATD3) \cite{WOS:001387362900037}, Multi-Agent Soft Actor-Critic (MASAC) \cite{WOS:001414873100010}, Cooperative Multi-Agent Reinforcement Learning (COMA) \cite{WOS:001483760900001}, and Multi-Agent Actor-attetion-Critic (MAAC) \cite{WOS:000684034303011} in determining the routing path.
Note that, to measure further the effectiveness of the interrupted policy based on MA-MAPPO, we also compare MA-MAPPO\_i with MA-MAPPO in ablation experiments.

To ensure the fairness of comparison, the compared algorithms' parameters are similar with our proposed algorithm as shown in Tab. \ref{table1}.
We carry out the evaluations from three aspects: 1) system convergence speed; 2) system routing effectiveness; 3) system availability and deployability.

\subsubsection{\textbf{System convergence speed}}
\begin{figure*}[t!]

  \subfloat[\label{fig5:a}]{
  \includegraphics[width=0.325\linewidth]{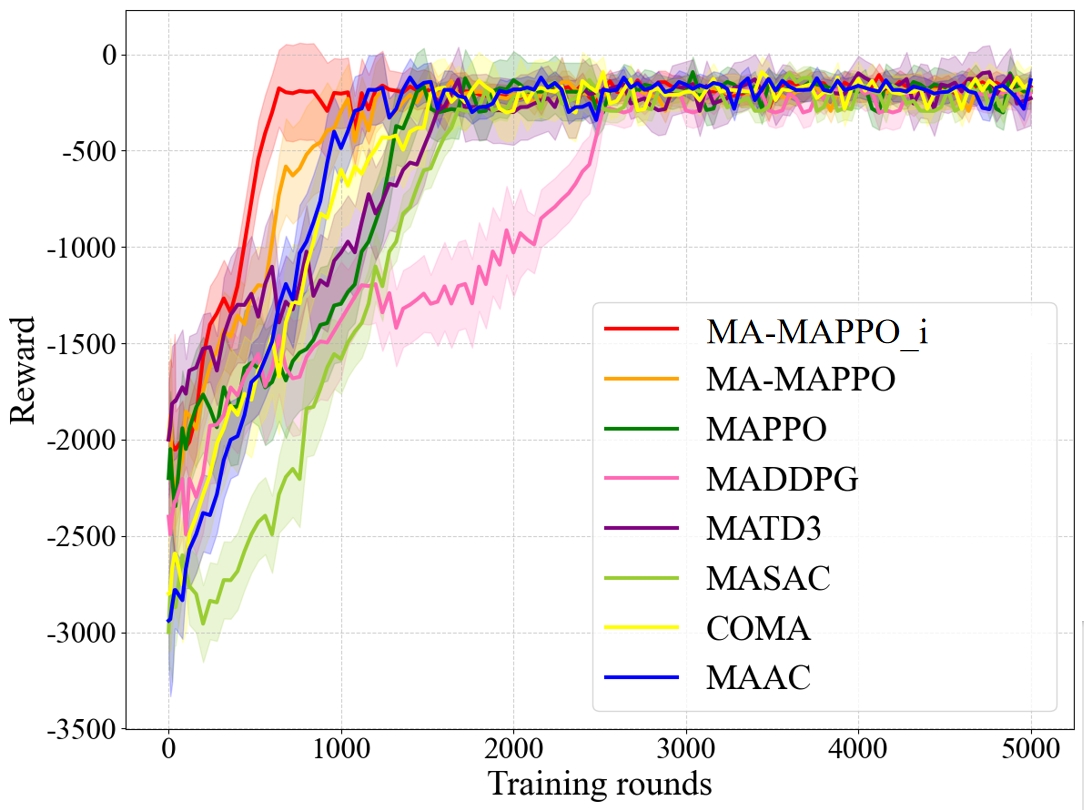}}
  \subfloat[\label{fig5:b}]{
  \includegraphics[width=0.325\linewidth]{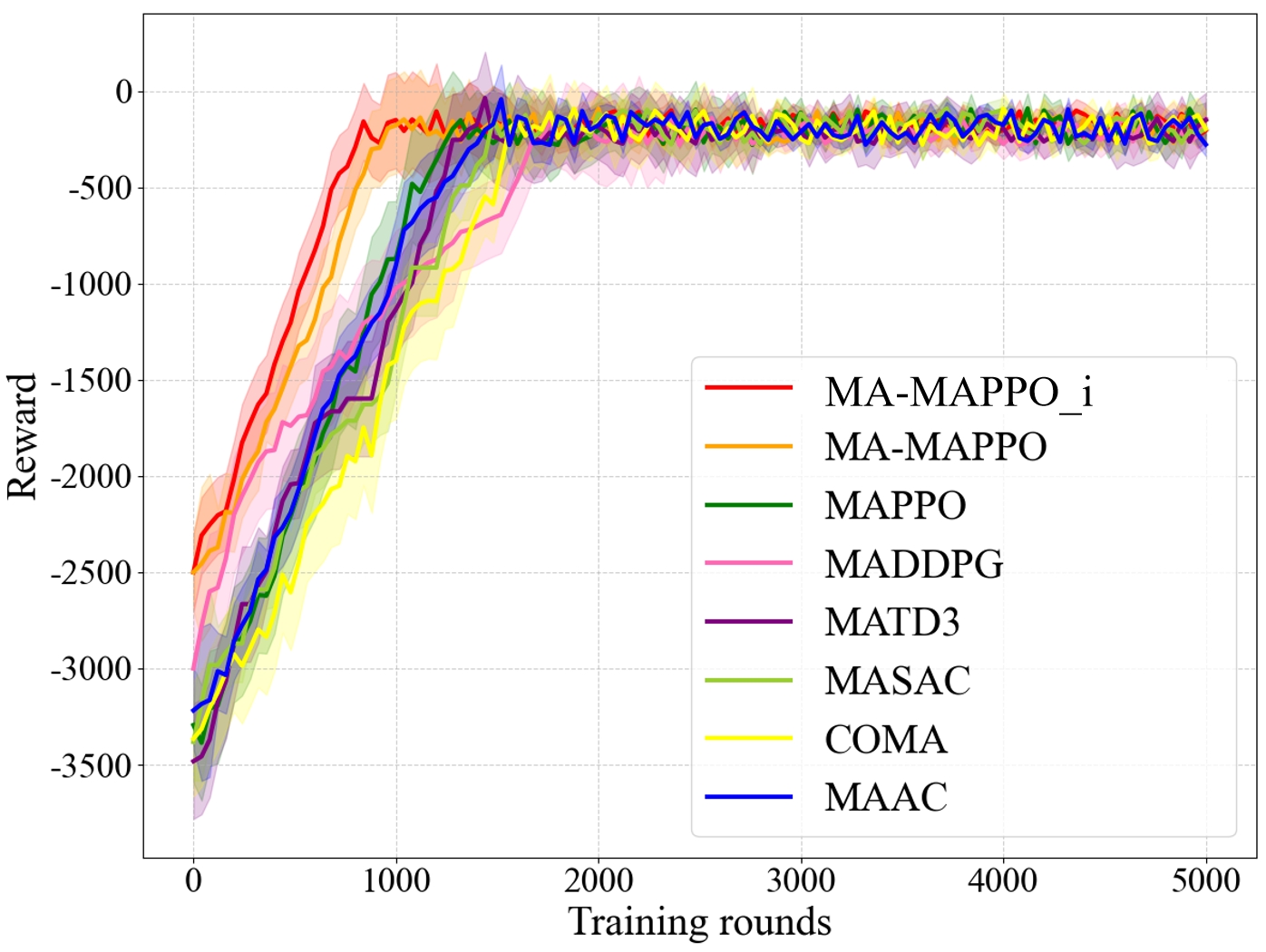}}
  \subfloat[\label{fig5:c}]{
  \includegraphics[width=0.325\linewidth]{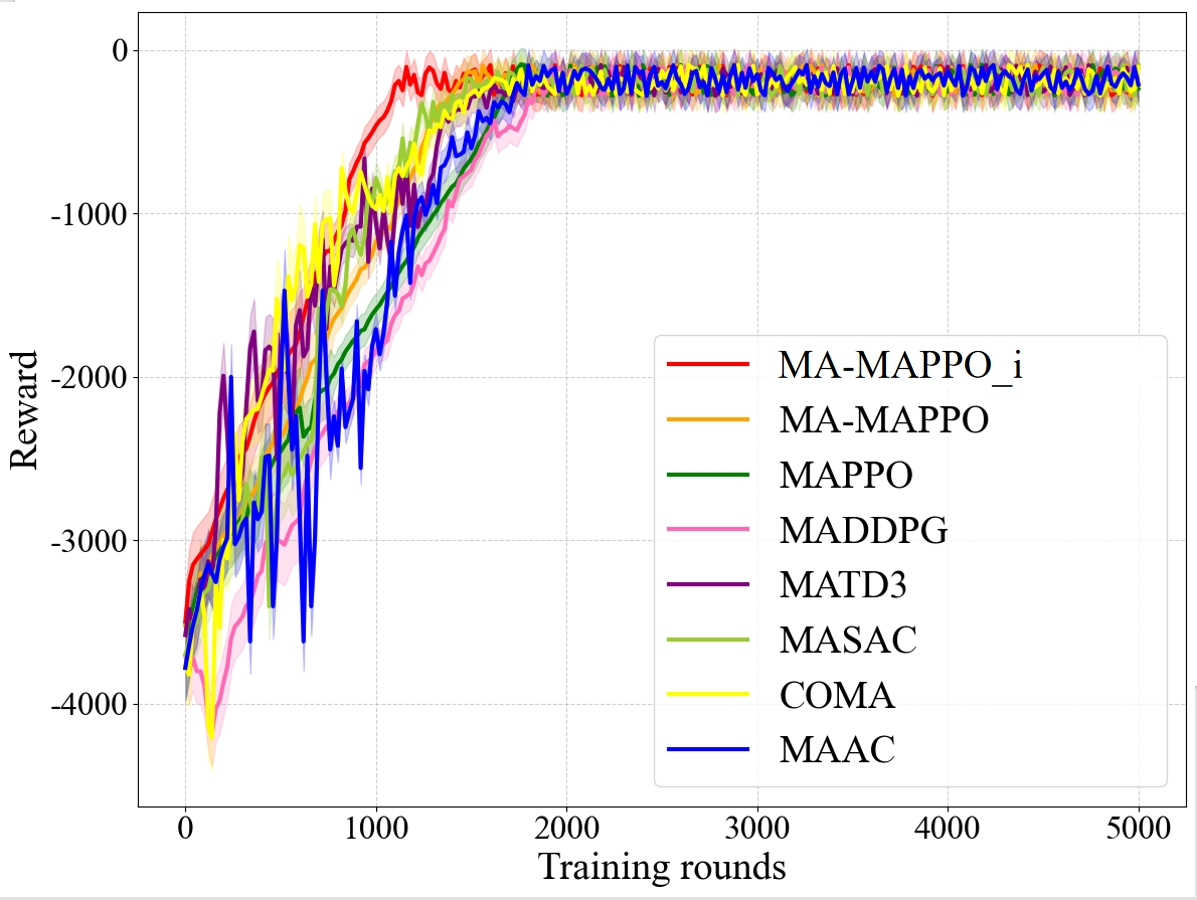}}

  \caption{System convergence speed comparison for UASNs with different numbers of nodes: (a) \(N_S = 64\), (b) \(N_S = 125\), (c) \(N_S = 216\).}
    \label{fig5}
\end{figure*}

To assess the convergence performance of the proposed MA-MAPPO\_i algorithm, we conduct multiple experiments across various routing scenarios. 
As displayed in Fig. \ref{fig5:a}, Fig. \ref{fig5:b}, and Fig. \ref{fig5:c}, MA-MAPPO\_i clearly outperforms the other algorithms. 
Especially in the early stages of training, MA-MAPPO\_i can rapidly converge to an optimal policy by dynamically adjusting the network view when nodes become unreachable, effectively responding to environmental changes and avoiding local optima.
The detailed observations and discussions are as follows:

The original MA-MAPPO uses a more conservative approach to policy updates, which limits its capacity to adapt to dynamic environments. 
In contrast, the traditional MAPPO algorithm, while exhibiting good convergence, updates its policy more conservatively, making it slower to adapt to changes in the environment, which leads to a tendency to converge to local optima in a complex environment. 
While MAPPO demonstrates favorable convergence characteristics in Fig. \ref{fig5}, its performance is suboptimal compared to MA-MAPPO, especially in environments with rapid changes.

Other algorithms, such as MADDPG, MATD3, and MASAC, show certain advantages in multi-agent collaborative tasks. 
However, these algorithms exhibit high computational complexity, particularly in dynamic and complex environments. 
MADDPG relies on deep reinforcement learning and utilizes deep neural networks to approximate Q-values, which implies the need to process more computations and memory, especially in high-dimensional state spaces, resulting in high computational costs and slower convergence. 
Similarly, MATD3 and MASAC suffer from comparable issues.
MATD3 incorporates a double delay mechanism and target networks to mitigate Q-value overestimation, but this increases computational complexity. 
MASAC, which integrates the soft actor-critic method, requires additional computation and optimization of both policy and value functions, further increasing its computational cost, particularly when multiple agents are involved in decision-making, leading to noticeable performance degradation.

In contrast, MA-MAPPO\_i significantly reduces unnecessary computations through the multi-head attention mask mechanism, enabling the algorithm to converge faster and more precisely. Additionally, the interrupted policy in MA-MAPPO\_i allows nodes to cease data transmission and dynamically update the network view when a node becomes unreachable, effectively addressing the uncertainties in multi-agent systems, reducing the burden of rapid environmental adaptation, and significantly improving convergence speed and stability. 
In summary, the results presented in Fig. \ref{fig5:a} - Fig. \ref{fig5:c} demonstrate that MA-MAPPO\_i exhibits superior adaptability and stability in complex environmental changes and multi-agent collaboration tasks.
By incorporating the interrupted policy and multi-head attention mask mechanism, MA-MAPPO\_i effectively handles environmental uncertainties while maintaining both exploration and convergence, significantly reducing the time and computational cost spent on ineffective exploration.

\begin{table*}[h!]
    \centering
    \renewcommand{\arraystretch}{1} 
    \footnotesize
    \caption{Mean propagation delay comparison}
    \begin{tabular}{lcccccccc}
        \toprule[1.5pt] 
        \(N_S\) & \textbf{MA-MAPPO\_i} & \textbf{MA-MAPPO} & \textbf{MAPPO} & \textbf{MADDPG} & \textbf{MATD3} & \textbf{MASAC} & \textbf{COMA} & \textbf{MAAC} \\
        \midrule
        64  & \textbf{9.21}  & 9.30 & 11.13 & 10.72 & 11.52 & 9.82 & 10.23 & 10.41 \\
        125 & \textbf{14.89} & 16.32 & 16.87 & 18.64 & 17.32 & 20.21 & 17.82 & 17.20 \\
        216 & \textbf{20.42} & 21.56 & 22.48 & 24.32 & 26.35 & 26.56  & 28.30 & 23.45 \\
        \bottomrule[1.5pt] 
    \end{tabular}
    \label{tab:2}
\end{table*}

\subsubsection{\textbf{System routing effectiveness}}
For the system performance evaluation, we utilize different algorithms to determine the routing path, especially in terms of delay and total routing execution time, respectively.
Note that in the evaluation scenario in this paper, to highlight the advantages of the proposed algorithm in routing path planning, only the propagation delay is considered.

\paragraph{Propagation delay}

\begin{figure*}[h!]
  \centering
  \subfloat[Delay comparison in \(N_S = 64\)\label{fig8:a}]{\includegraphics[width=0.3\textwidth]{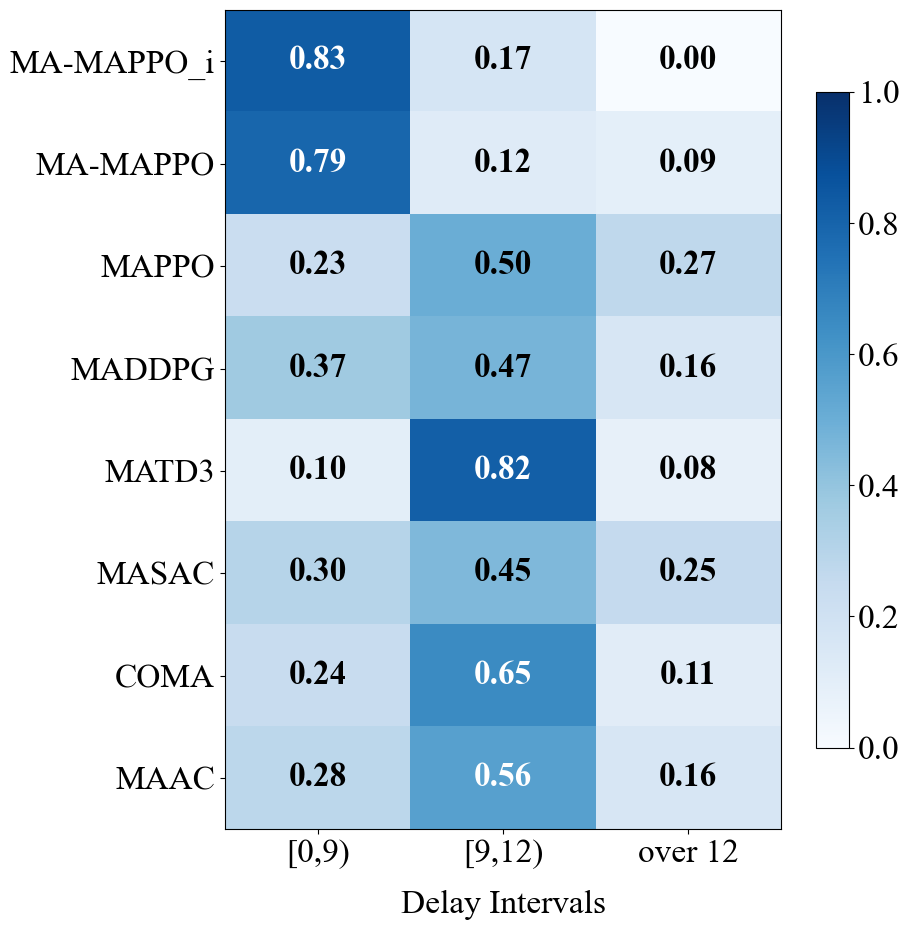}}\hfill
  \subfloat[Delay comparison in \(N_S = 125\)\label{fig8:b}]{\includegraphics[width=0.3\textwidth]{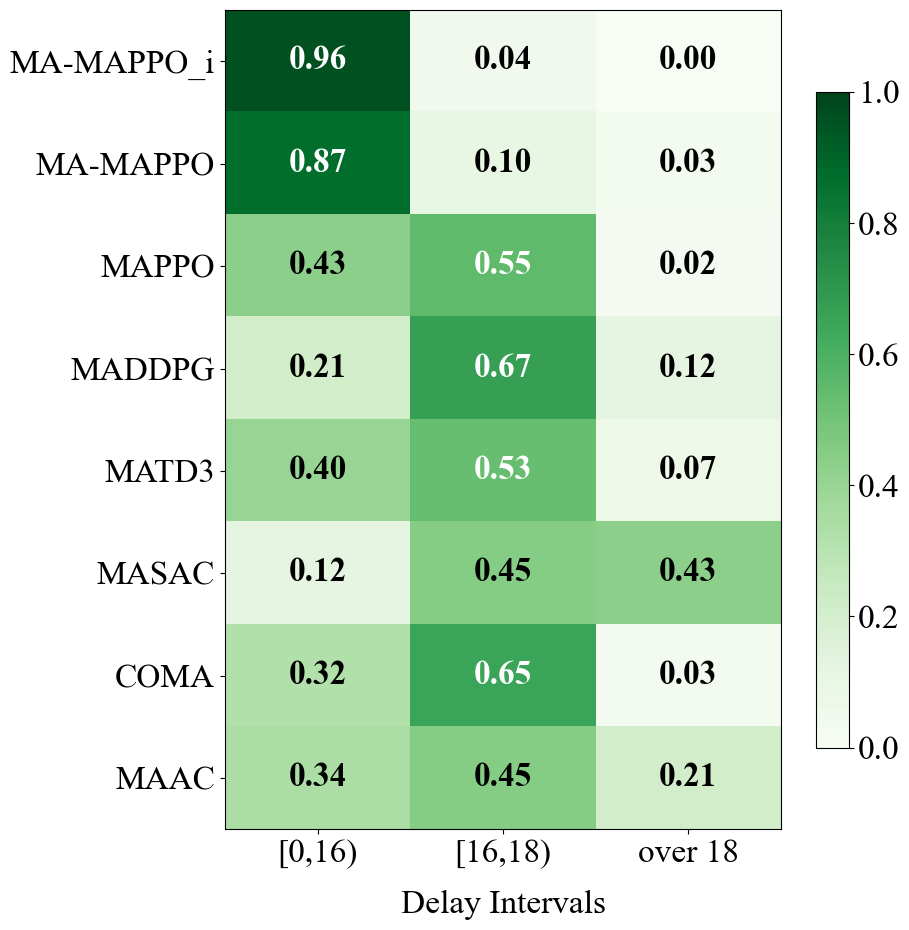}}\hfill
  \subfloat[[Delay comparison in \(N_S = 216\)\label{fig8:c}]{\includegraphics[width=0.3\textwidth]{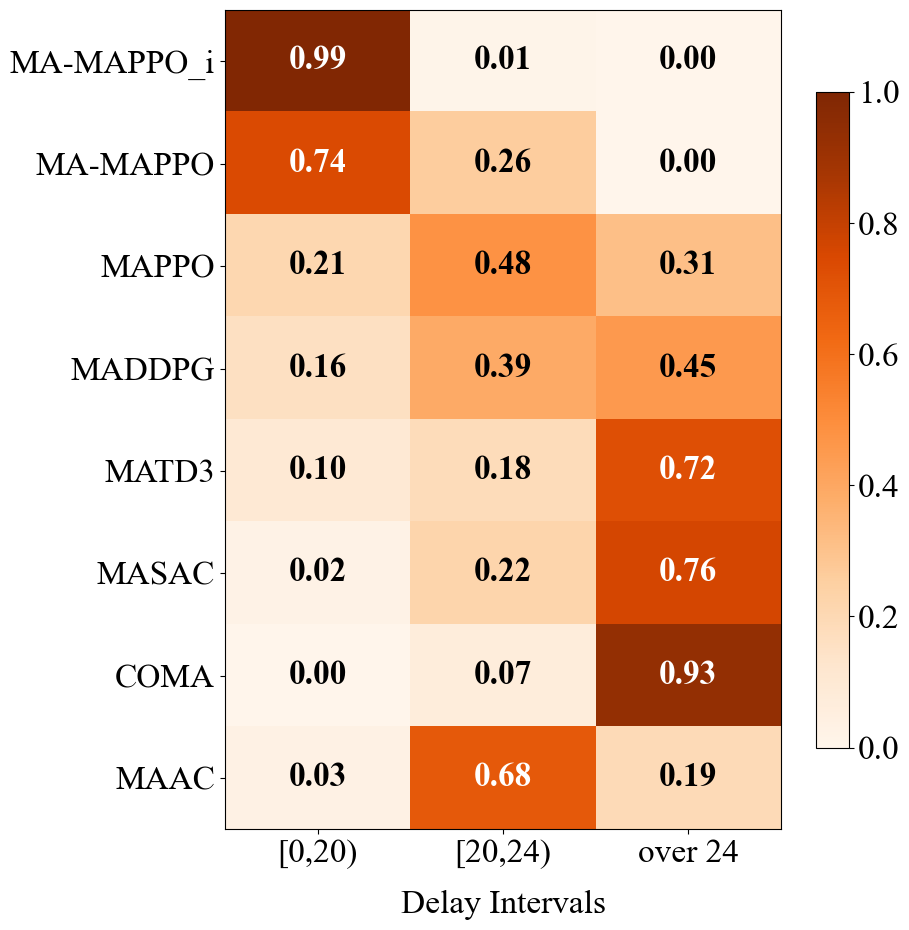}}\hfill
  \caption{Distribution for propagation delay comparison across different numbers of nodes: (a) \(N_S = 64\) with the intervals of [0, 9), [9, 12), and over 12, (b) \(N_S = 125\) with the intervals of [0,16), [16, 18), and over 18, (c) \(N_S = 216\) with the intervals of [0, 20), [20, 24), and over 24.}
  \label{fig8}
\end{figure*}

Propagation delay tests the time taken for a packet to travel from the source node to the sink node.
The propagation of a routing path for packet \(p\) with \(H\) hops can be computed as Eq. \ref{EQdelay}:
\begin{equation}
\label{EQdelay}
    T_{delay}(p)=\sum^H_{p=1} \frac{D_p}{v_s},
\end{equation}
where \(T_{delay}(p)\) represents the propagation delay of each packet, \(D_p\) is the Euclidean distance between consecutive nodes on the path, and \(v_s\) represents the speed of sound in water (default setting to 1531 m/s, adjusted with the environment).
The results in Fig. \ref{fig8} and Tab. \ref{tab:2} collectively illustrate the superior performance of MA-MAPPO\_i and MA-MAPPO.
As seen in Tab. \ref{tab:2}, for $N_S = 64$, MA-MAPPO\_i has a delay of 9.21, far lower than other algorithms like MAPPO (11.13). 
In Fig. \ref{fig8:a}, it also dominates the low-delay interval $[0,9)$ with a proportion of 0.83, showcasing its higher efficiency.  
MA-MAPPO builds on the MAPPO framework, integrating multi-head attention mask mechanism. 
These allow it to filter out invalid action spaces, simplifying input and reducing redundant computations. 
MA-MAPPO\_i further introduces an interrupted policy.
By which, if a next-hop node becomes unreachable (e.g., due to energy depletion), the current node halts data transmission, stores packets, and requests the CA sensor node to update the network view. 
This dynamic response minimizes unnecessary delays.

In contrast, traditional algorithms such as MAPPO, MADDPG, and others exhibit suboptimal performance in delay. 
MAPPO, lacking targeted mechanisms for invalid action filtering, struggles with complex input spaces. 
As $N_S$ increases (e.g., to 216), the absence of effective input simplification leads to more redundant transmissions, resulting in higher delays (Tab. \ref{tab:2} shows MAPPO’s delay at 22.48 for $N_S = 216$).
MADDPG relies on centralized training, which often accumulates delays in a dynamic underwater environment. 
Algorithms like MATD3 fail to balance exploration and exploitation effectively, making it hard to converge to stable strategies quickly. 
This is also reflected in Fig. \ref{fig8:a} - Fig. \ref{fig8:c}, where they have higher proportions in medium-to-high delay intervals, unlike the MA-MAPPO or MA-MAPPO\_i. 

In particular, as $N_S$ rises from 64 to 216, network complexity grows exponentially. 
MA-MAPPO\_i maintains its edge via the interrupted policy. 
When node failures become more frequent with increased $N_S$, it promptly pauses invalid transmissions and updates the network view, keeping delays low. 
For $N_S = 216$, its delay in Tab. \ref{tab:2} is 20.42, and in Fig. \ref{fig8:c}, it dominates the $[0,20)$ interval with a proportion of 0.99. 
Note that, even if MA-MAPPO is not equipped with the interrupted policy, it still benefits from attention and masking to simplify input, outperforming traditional algorithms. 
However, without a dynamic failure response, MA-MAPPO's delay (21.56 for $N_S = 216$) is slightly higher than MA-MAPPO\_i.
Due to poor input handling and adaptation, the delays of traditional algorithms surge as $N_S$ increases, highlighting the adaptability of MA-MAPPO and MA-MAPPO\_i.

\paragraph{Total execution time}

To further measure the routing effectiveness of the proposed approaches, we evaluate the total execution time \(T_e\) it takes to complete the routing task, which can be expressed by Eq. \ref{eqexe}:
\begin{equation}
    \label{eqexe}
    T_e=\frac{N_D\cdot S}{B}+\sum_{p=1}^{N_D} T_{delay}(p),
\end{equation}
where \(T_e\) represents the total execution time, \(S\) represents the average packet size, \(B\) represents the available bandwidth.
Testing the total execution time can verify the utilization of system resources (e.g., bandwidth constraints) and the adaptation to dynamic network requirements. 
For networks of different scales, we preset distinct numbers of pending packets ($N_D$). 

\begin{figure*}[h!]
  \centering
  \includegraphics[width=0.8\linewidth]{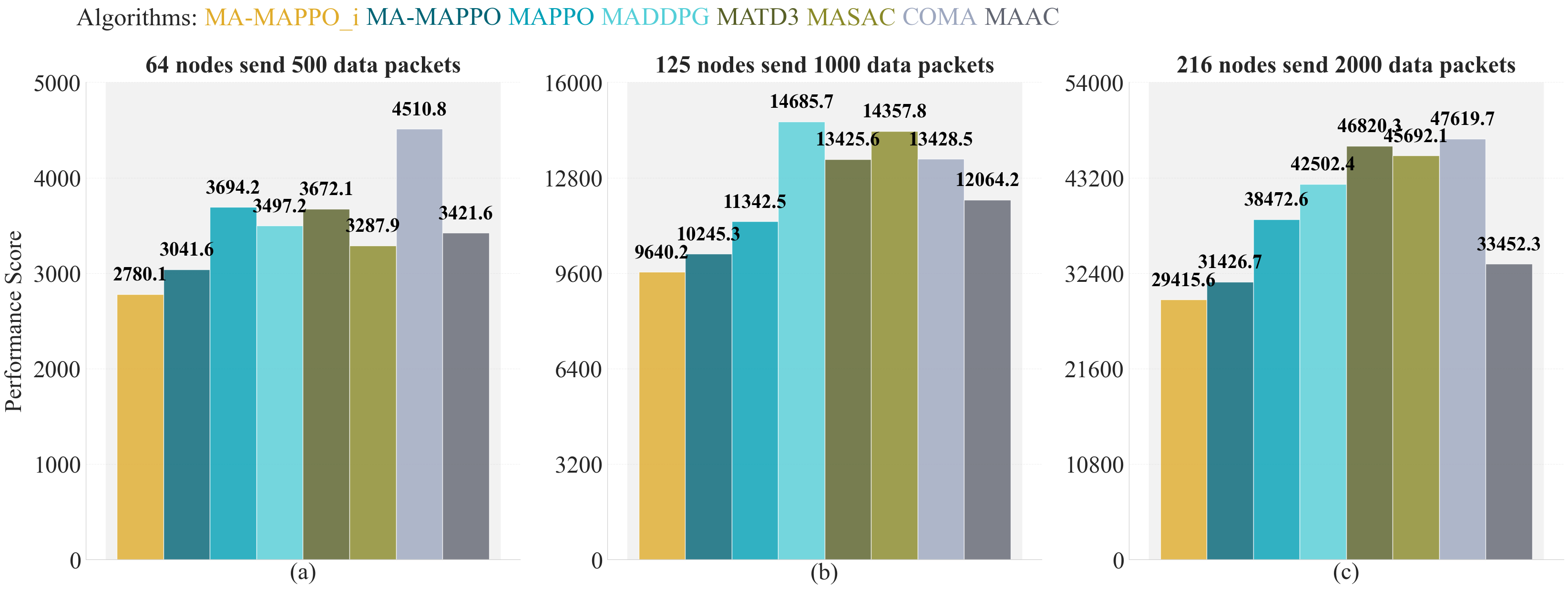}
  \caption{Total execution time comparison for different scales of networks: (a) 500 data packets to be forwarded in the network with scale of \(N_S=64\), (b) 1000 data packets to be forwarded in the network with scale of \(N_S=125\), (c) 2000 data packets to be forwarded in the network with scale of \(N_S=216\).}
  \label{fig9}
\end{figure*}

The results in Fig. \ref{fig9} indicate that the proposed MA-MAPPO\_i shows the best performance.
Even if MA-MAPPO\_i maintains a lower total execution time, as the number of sensor nodes and data packets increases.
As aforementioned, MA-MAPPO\_i introduces the multi-head mask mechanism to shield invalid action spaces, simplifying the input. 
With the interrupted policy, when the next-hop node is unreachable, the current node stops sending, stores packets, and requests CA sensor node to update the network view. 
This policy can help reduce unnecessary transmission attempts, thus the execution time can be saved.

By comparison, the MAPPO, MADDPG, MATD3, etc., naturally underperform compared to MA-MAPPO or MA-MAPPO\_i in relevant aspects.
These result from the fact that the compared algorithms are not equipped with multi-head attention mask mechanism and the interrupted policy, may engage in more redundant actions in dynamic network situations. 

Totally, the proposed MA-MAPPO\_i can effectively balance resource utilization during policy training, adapting to dynamic network demands.
This allows it to handle scalability better, avoiding excessive time consumption from redundant computations or inappropriate transmissions.

\begin{figure}[h!]
  \centering
  \includegraphics[width=1\linewidth]{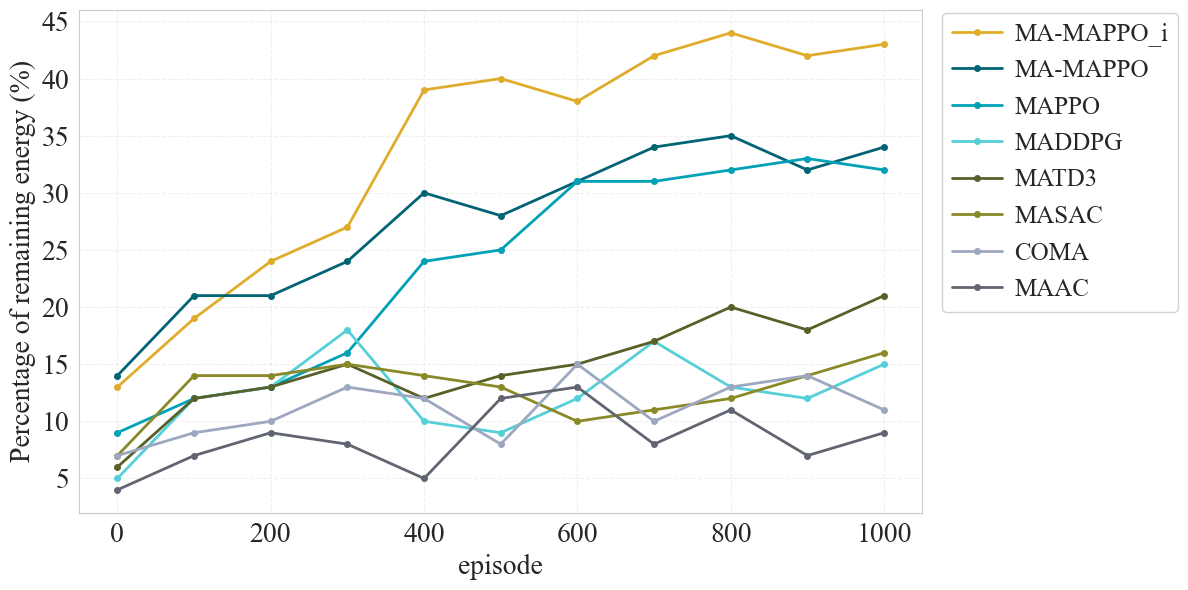}
  \caption{Test for remaining energy in \(N_S = 216\) }
  \label{fig10}
\end{figure}

\paragraph{Energy consumption}
To visually demonstrate the improvements of MA-MAPPO\_i brings to the energy saving in the routing process, we demonstrate the test for the remaining energy is shown in Fig. \ref{fig10}. We model the energy consumption process in Eq. \ref{EQenergy} \cite{WOS:001198016900017}.

\begin{equation}
\label{EQenergy}
    P_{rem} = \frac{E_{init} - \sum_{i=1}^{N_{tx}} E_{tx}^{(i)} - \sum_{j=1}^{N_{rx}} E_{rx}^{(j)}}{E_{init}},
\end{equation}
where $P_{rem}$ represents the percentage of remaining energy, \(N_{tx}\) and \(N_{rx}\) represent the total times of transmitting or receiving data packets. \(E_{tx}\) and \(E_{rx}\) represent the energy consumption of transmitting or receiving a data packet, respectively.

As illustrated in Fig. \ref{fig10}, the proposed MA-MAPPO\_i achieves superior energy efficiency by integrating the interrupted policy with a multi-head attention mask mechanism.
MA-MAPPO\_i dynamically suspends transmissions upon next-hop node unavailability, avoiding wasted energy from repeated attempts, while its attention mechanism pre-filters invalid actions to reduce redundant exploration.
In comparison, MA-MAPPO employs attention-based masking but lacks real-time interruption handling, leading to higher energy waste during link failures \cite{WOS:001269813700001}. 
Traditional algorithms like MAPPO and MADDPG perform even poorer, while MAPPO uses conservative policy updates without action filtering, and MADDPG suffers from Q-value overestimation due to centralized training.
Both of MAPPO and MADDPG lack rapid response to interruption caused by node failures, causing persistent energy drain on unreachable paths.
Thus, MA-MAPPO\_i’s quick recovery to interruption ensures minimal energy waste in highly dynamic UASNs.

\subsubsection{\textbf{System availability and deployability}}

Ultimately, to clearly demonstrate the availability and deployability of our entire routing process directed by our proposed MA-MAPPO\_i based on ISURL, we construct an underwater environment based on the GEBCO ocean dataset \cite{WOS:001167083800025}.

\begin{figure*}[t!]
  \centering
  \subfloat[3D scenario for routing\label{fig11:a}]{\includegraphics[width=0.5\textwidth]{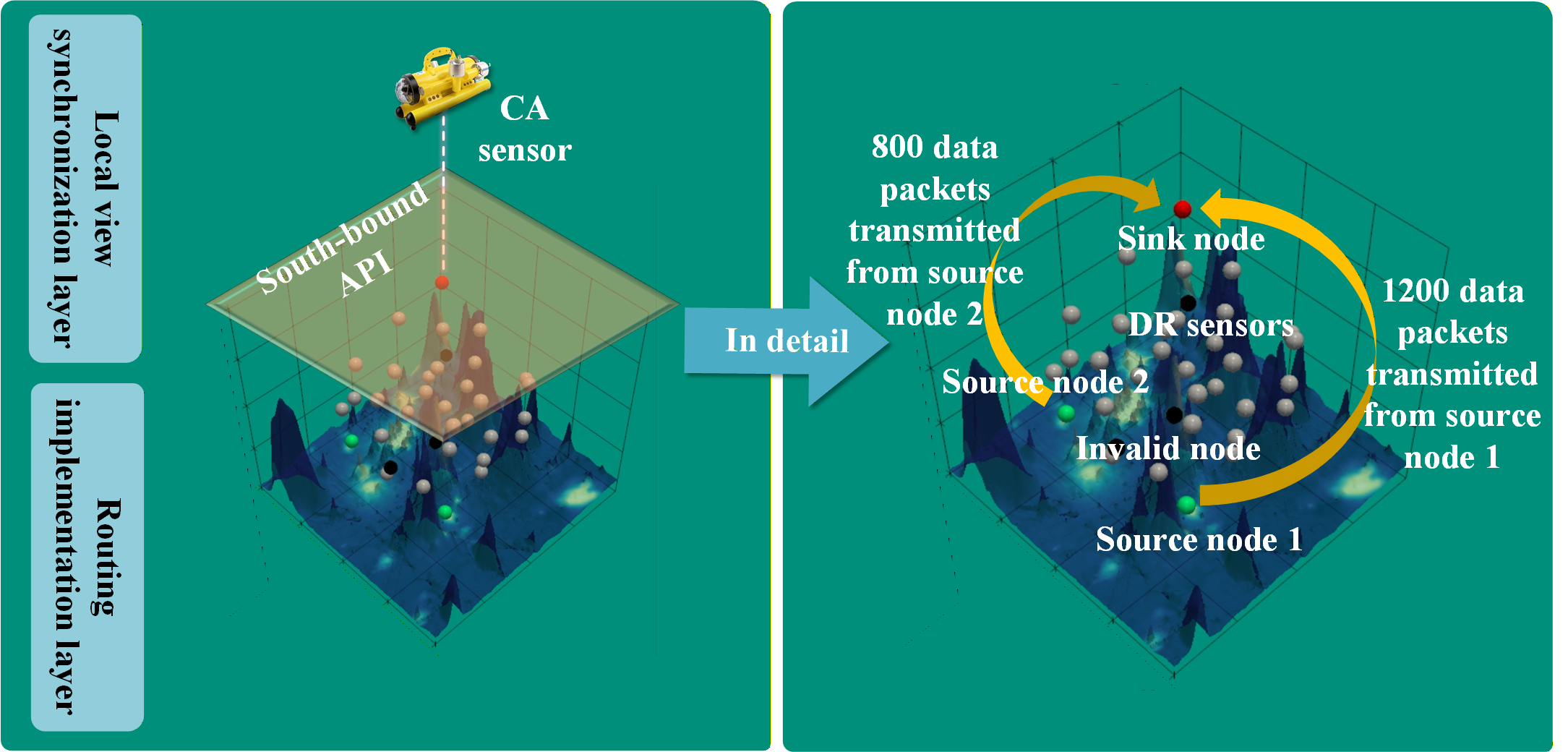}}\hfill
  \subfloat[Real routing transmission paths and trained routing transmission from the top view for source node 1\label{fig11:b}]{\includegraphics[width=0.24\textwidth]{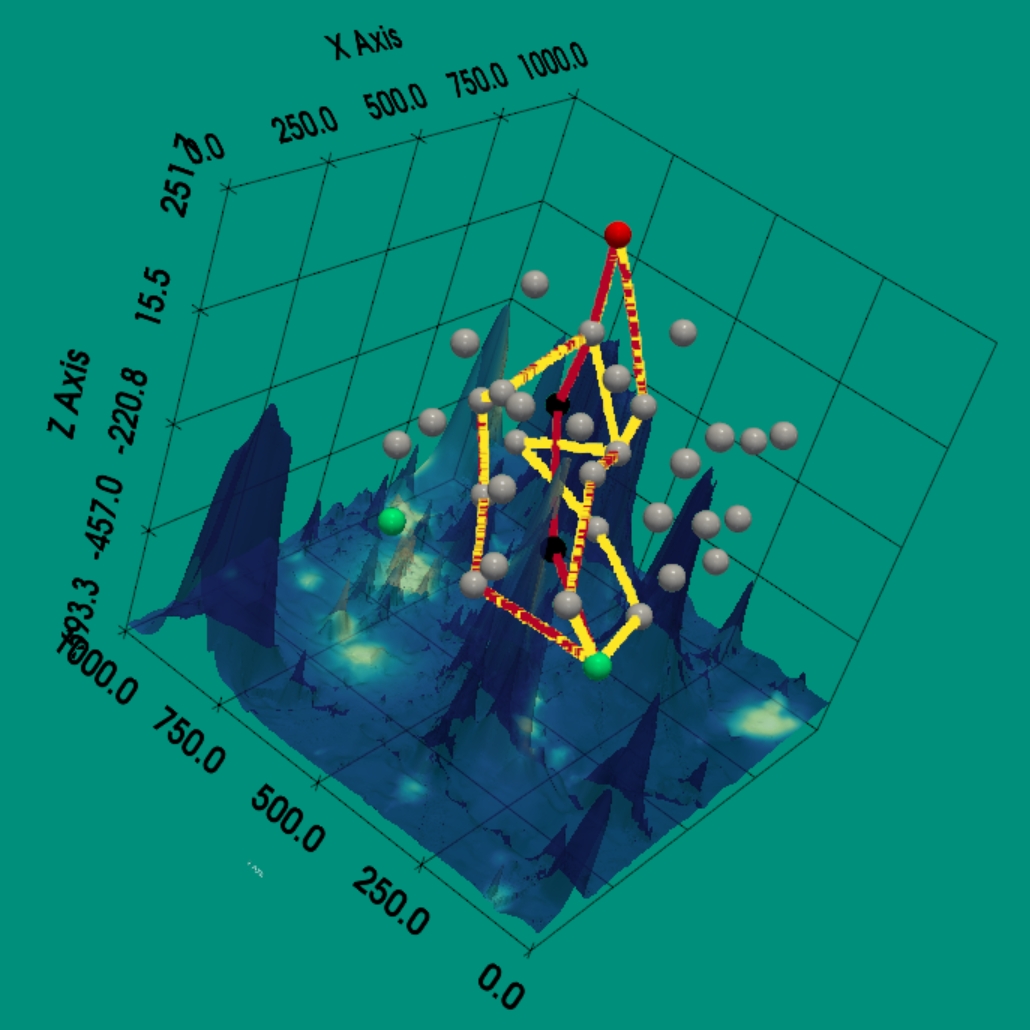}}\hfill  
  \subfloat[Trained routing paths for source node 1\label{fig11:c}]{\includegraphics[width=0.24\textwidth]{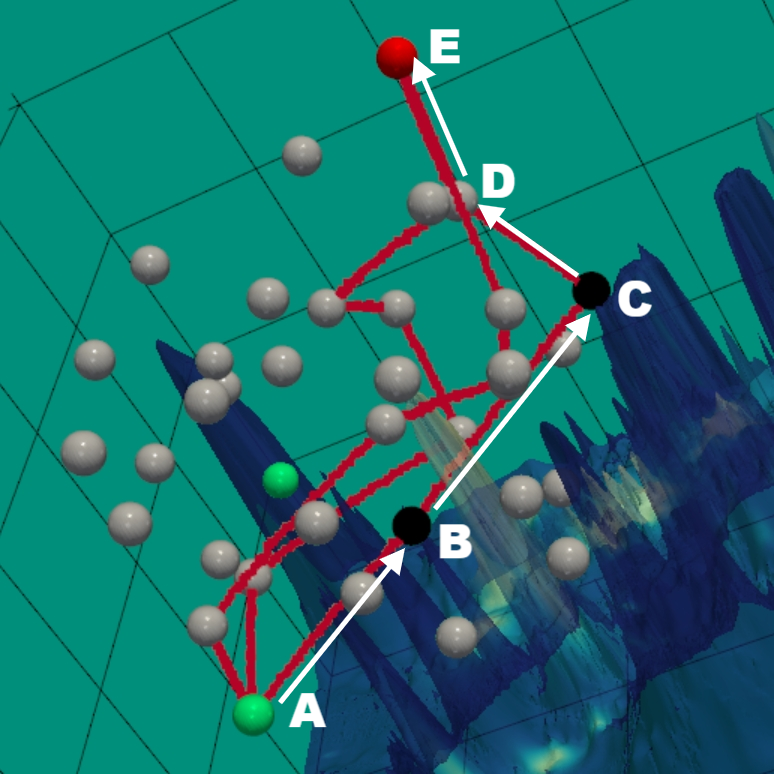}}\hfill
  
  \subfloat[Real routing paths for source node 1\label{fig11:d}]{\includegraphics[width=0.24\textwidth]{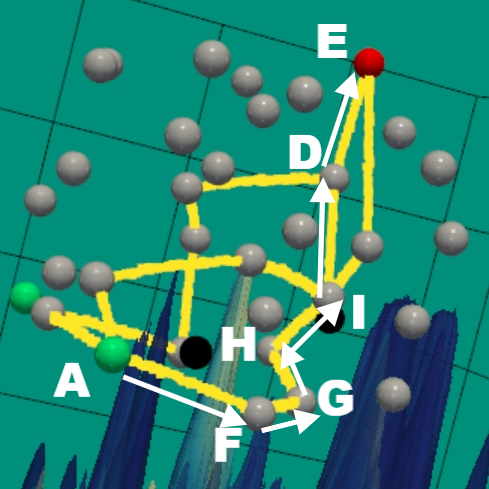}}\hfill
  \subfloat[Real routing transmission paths and trained routing transmission from the top view for source node 2\label{fig11:e}]{\includegraphics[width=0.24\textwidth]{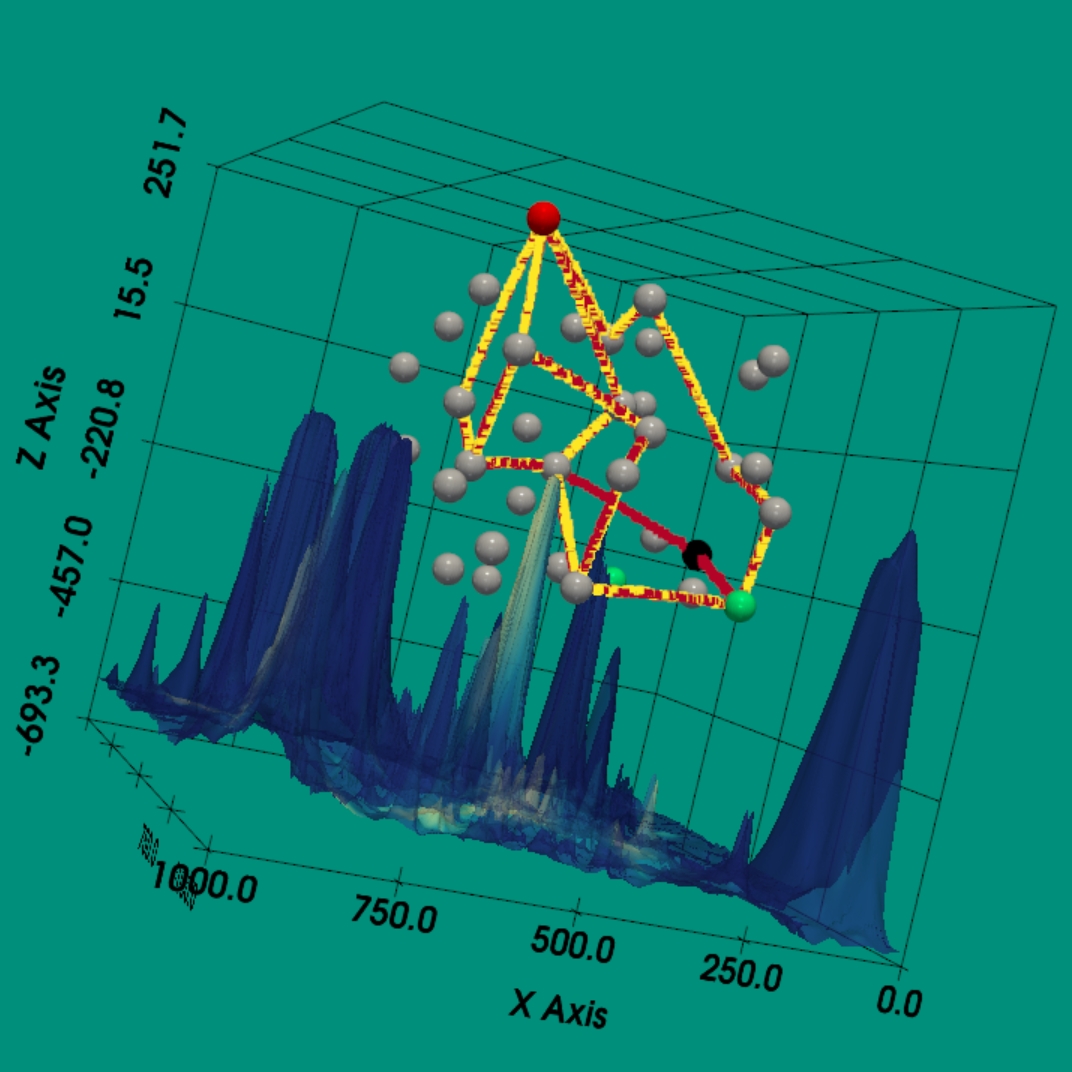}}\hfill
  \subfloat[Trained routing paths for source node 2\label{fig11:f}]{\includegraphics[width=0.24\textwidth]{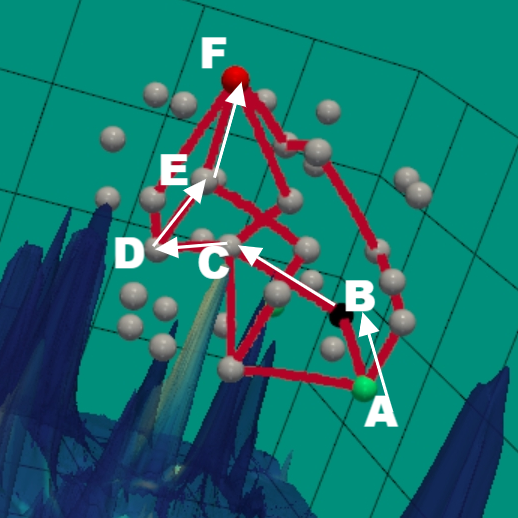}}\hfill
  \subfloat[Real routing paths for source node 2\label{fig11:g}]{\includegraphics[width=0.24\textwidth]{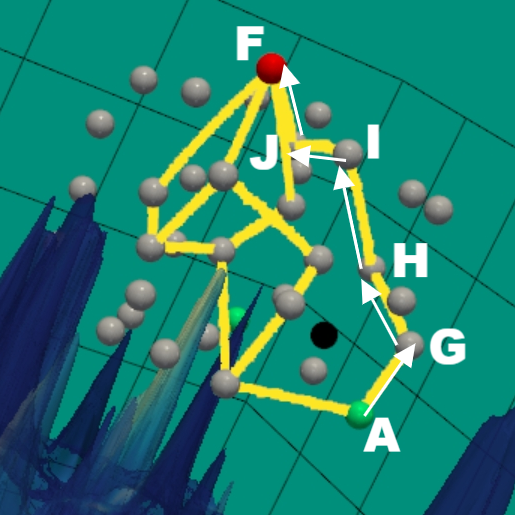}}\hfill
  
  \caption{3D scenario for routing availability and deployability test. (a) shows the 3D scenario of the model with selected data of the latitude range of 3.38°N-16.63°N and the longitude range of 157.37°E-161.41°E of the oceanic environment, and its relevant data are selected for the system model to visualize the routing process. (b) - (d) shows the routing decision procedure from source node 1. (e) - (g) displays the routing decision procedure from source node 2. The yellow curve represents the actual routing execution paths, while the red curve represents the routing paths trained by the proposed MA-MAPPO\_i. Note that if the actual routing paths coincide with the trained routing paths, the red curve will overlap with the yellow curve, as shown in (b) and (e). }  
  \label{fig11}
\end{figure*}

Using a simple demonstration, we select 36 nodes (3x3x4 grid-size) from a randomly generated network of 216 nodes (the parameters are similar with what we did in Fig. \ref{fig5} - Fig. \ref{fig9}), to showcase the availability and deployability of the proposed method.
Fig. \ref{fig11:a} illustrates the 3D scenario for the availability and deployability test, where the active data routing is triggered from source node 1 and source node 2 to a common sink node. 
In Fig. \ref{fig11:a}, the spheres represent DR nodes in the routing process, noting that the green nodes represent the source node, the red node represents the sink node and the black nodes represent the invalid node (which cannot perform routing tasks).

For the routing triggered from source node 1, as shown in Fig. \ref{fig11:b}, among the selected 36 nodes during the training, 13 nodes are chosen by MA-MAPPO\_i as the routing forwarding nodes (nodes along the routing path), and 14 nodes participate in the actual routing process. 
We select one specific path for detailed analysis. 
As shown in Fig. \ref{fig11:c}, during the training phase, the MA-MAPPO\_i algorithm optimizes the ideal path $<A, B, C, D, E>$ based on the node states and environmental conditions. 
However, in the actual execution process, the trained routing path is not adopted, and the actual routing path is switched to $<A, F, G, H, I, D, E>$.
This discrepancy can be attributed to the dynamic environmental changes in the actual routing execution.
According to the ISURL framework proposed in Sec. \ref{Section:4}, the DR node detects that the next-hop node in the routing path is unreachable (for example, node B in Fig. \ref{fig11:c}.
Due to node B failure or a change in the distance between nodes A and B, it prevents the establishment of a valid routing path and results in the interruption of the routing procedure, making the DR node first temporarily storage the current data packet (the cash size \(C_{size}\) of each node is defined in Tab. \ref{table1}).
After temporarily storing the data packet, the DR sensor node reports this issue to the upper-level CA sensor.
Upon receiving the report, the CA sensor uses the routing mechanism based on the MA-MAPPO\_i algorithm and the mask function to assess the reachability of all neighboring nodes under its management.
The CA sensor then compiles the reachability information from these nodes into a global network topology. 
Based on this global topology, the CA sensor retrains the routing strategy to optimize the routing paths. 
Subsequently, the DR sensor will forward data based on the newly optimized routing strategy which is shown in Fig. \ref{fig11:d}, to ensure that the data can be transmitted through the newly accessible path.

Fig. \ref{fig11:e} - Fig. \ref{fig11:g} illustrate the similar routing processing with Fig. \ref{fig11:b} - Fig. \ref{fig11:d}, where the routing paths trained by the MA-MAPPO\_i algorithm is $<A, B, C, D, E, F>$ in Fig. \ref{fig11:f}, but the actual routing path during execution is $<A, G, H, I, J, F>$ as shown in Fig. \ref{fig11:g}. 
The reason for this discrepancy is that an effective routing path could not be established between nodes A and B, causing a routing interruption. 
As a result, the MA-MAPPO\_i-based routing mechanism updates the network view and retrains the routing strategy based on the updated view.

Therefore, the difference between the actual routing path and the trained path reflects the impact of dynamic environments and node failures on the routing process. 
Note that our proposed ISURL framework allows for dynamic adjustment of the routing strategy to adapt to these environmental changes.

\begin{figure}[h!]
  \centering
  \subfloat[The tested edge platform\label{fig13:a}]
    {\includegraphics[width=0.2\textwidth]{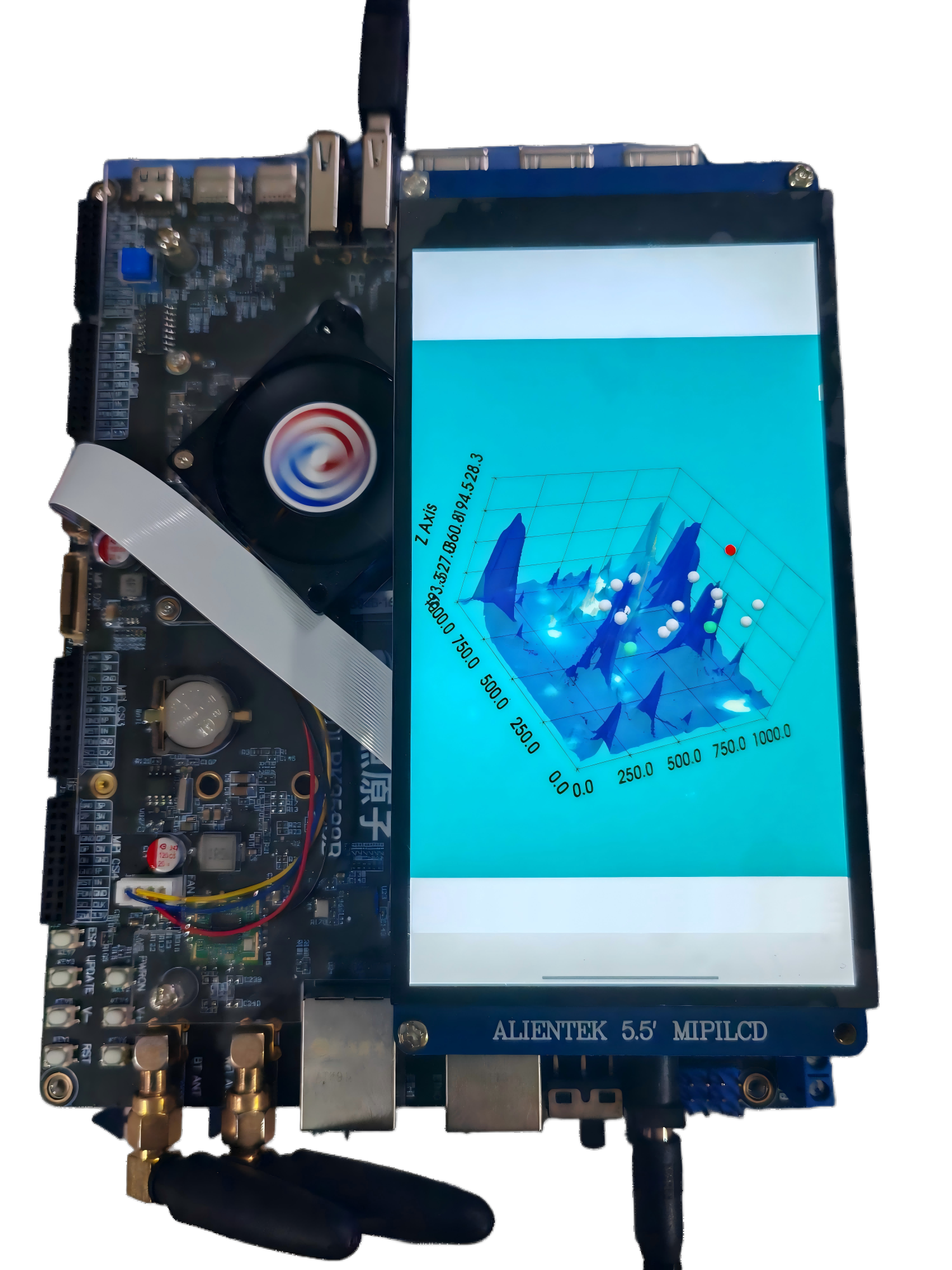}}\hfill
  \subfloat[Network view\label{fig13:b}]
    {\includegraphics[width=0.28\textwidth]{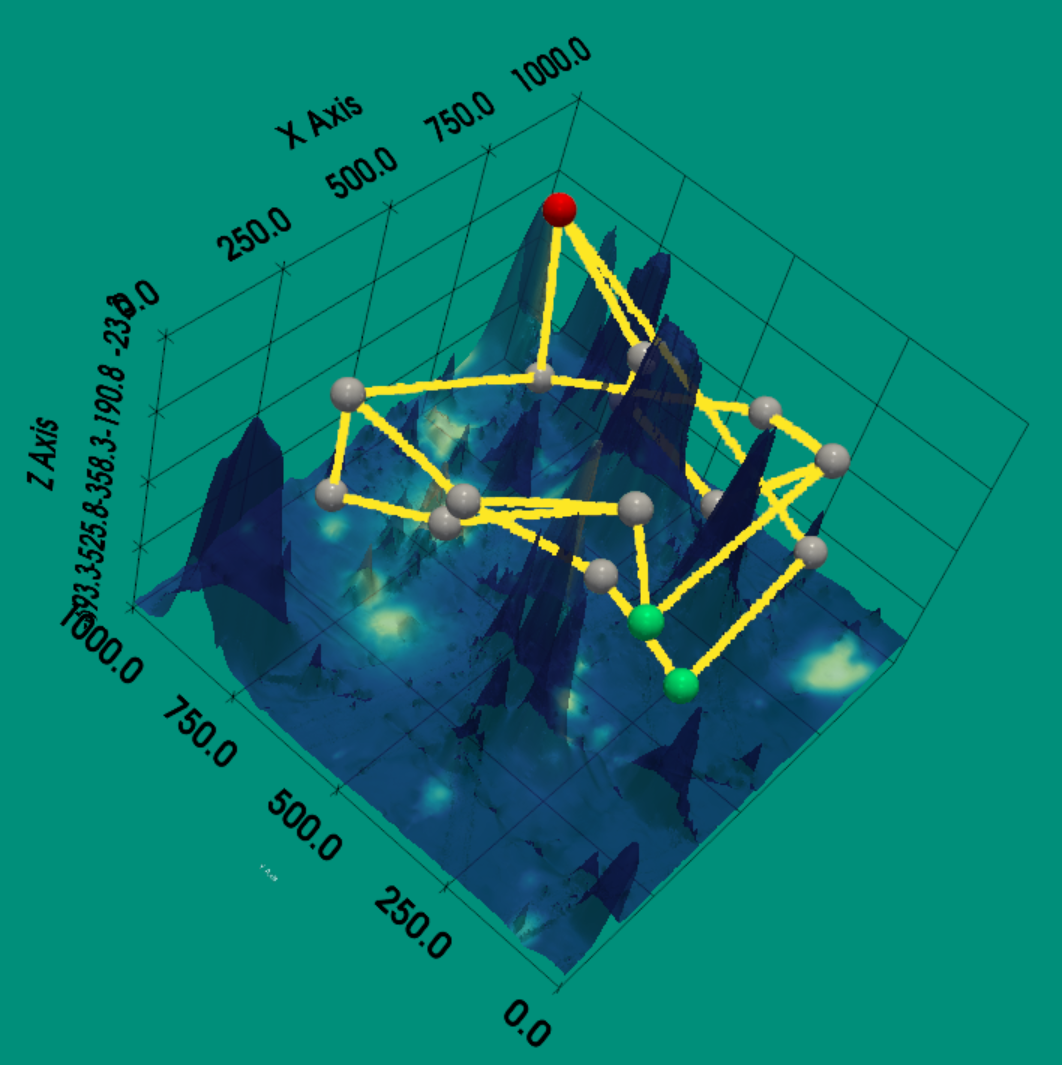}}
  \caption{Test on an edge platform}
  \label{fig13}
\end{figure}

Moreover, to further test the availability and deployability of the proposed routing scheme based on the MA-MAPPO\_i, we deploy the model on the RK3588 edge platform.
Fig. \ref{fig13:a} shows the edge platform we use, and on its screen we deploy an offline program to see the routing process of the network.
Fig. \ref{fig13:b} shows the network view deployed on the edge platform, specifically, the green nodes represent the source nodes and the red node represents the sink node.

  

\section{Conclusion}\label{Section:7}

In this paper, we focus on smart data routing for UASNs. 
Leveraging the SDN technique and interrupted policy, we present ISURL, an interrupted software-defined UASNs MARL framework that divides the network into three functional layers and performs MARL-driven routing decision-making.
To facilitate smart routing decisions in the complex underwater environment, we propose an autonomous routing decision algorithm based on a proposed MA-MAPPO which improves the MAPPO with multi-head attention mask mechanism. 
Furthermore, on account of the influences of the invalid routing paths caused by corrupted nodes or farther physical distance between sensor nodes, we introduce an interrupted policy to MA-MAPPO and propose MA-MAPPO\_i, which allows for more flexible handling of invalid routing paths and prevents the emergent routing failure.
The MA-MAPPO\_i algorithm significantly reduces the input space dimensions to simplify the training process by filtering out invalid action space.
The evaluation results showcase that the proposed routing scheme can quickly identify optimal routing paths in UASNs and achieve more efficient data routing compared to several mainstream approaches.

The future research directions derived from this work include: 
1) exploring energy-efficient routing algorithms for UASNs that consider the energy constraints of sensor nodes and optimize routing decisions to prolong network lifetime, particularly in energy-harvesting underwater environments;
2) investigating the integration of machine learning and sensor fusion techniques to enhance real-time decision-making in dynamic underwater environments, allowing the system to adapt to sudden environmental changes or network topology variations;
3) advancing the deployment strategies on hardware for MARL-driven routing protocols, with a focus on embedding trained models into resource-constrained underwater devices (e.g., ARM-based platforms). 
This involves refining the scale of action-state spaces by incorporating low-level hardware metrics (e.g., CPU utilization, memory footprint, and energy consumption per computation cycle) into the MARL framework.
Additionally, we will investigate federated learning mechanisms to enable distributed, on-device training and offline execution while minimizing communication overhead, ensuring the system’s scalability and robustness in large-scale UASNs deployments under stringent resource conditions.

\appendices


\ifCLASSOPTIONcaptionsoff
  \newpage
\fi



%
\bibliographystyle{IEEEtran}
\bibliography{ref}

\vspace{-5ex}

\begin{IEEEbiography}[{\includegraphics[width=1in,height=1.25in,clip,keepaspectratio]{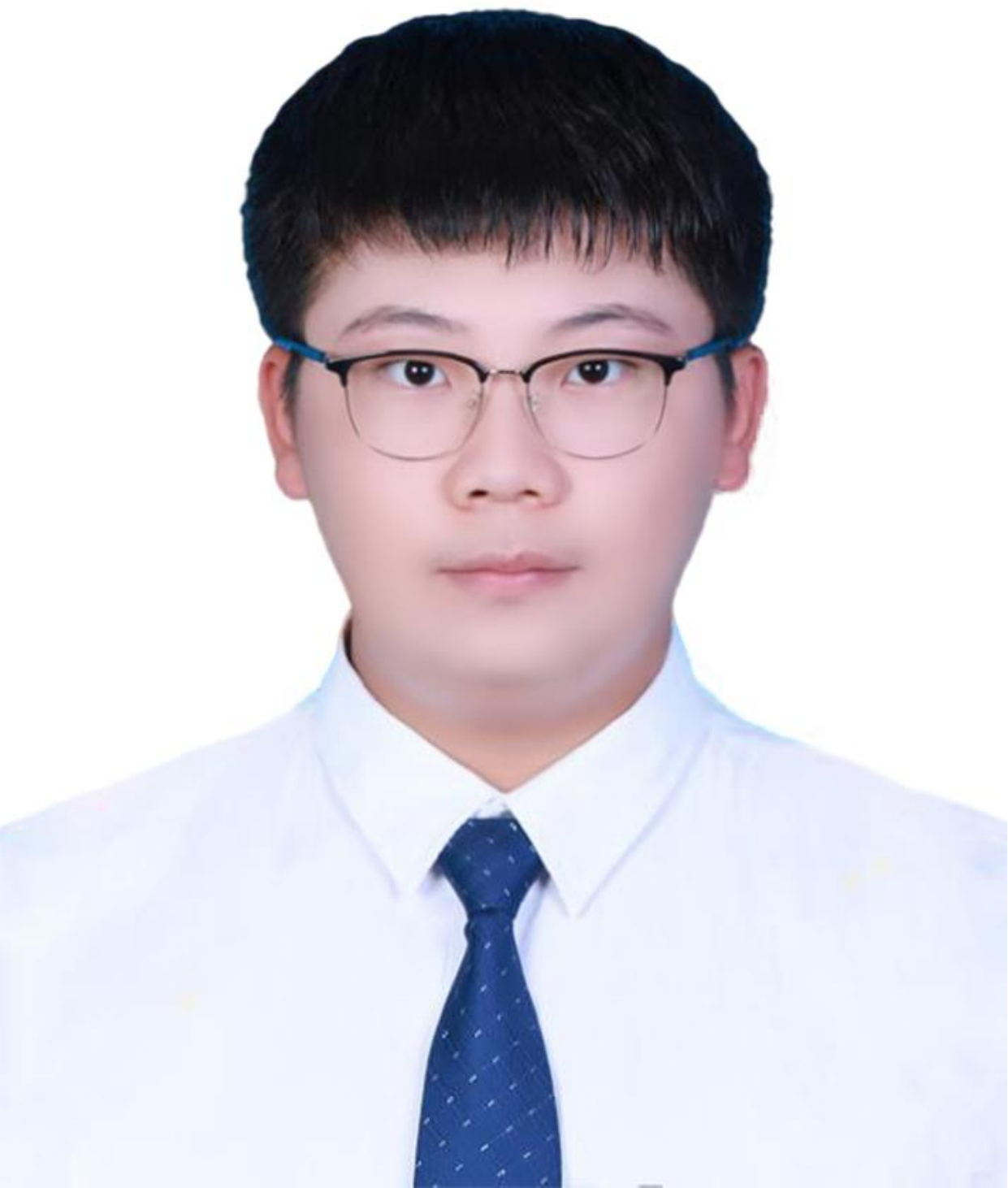}}]{Zhenyu Wang}
(Student member, IEEE) is currently pursuing a Bachelor's degree at the Software College, Northeastern University, Shenyang, China. His research interests include reinforcement learning, Internet of things and software-defined networking. 
\end{IEEEbiography}

\vspace{-5ex}
\begin{IEEEbiography}[{\includegraphics[width=1in,height=1.25in,clip,keepaspectratio]{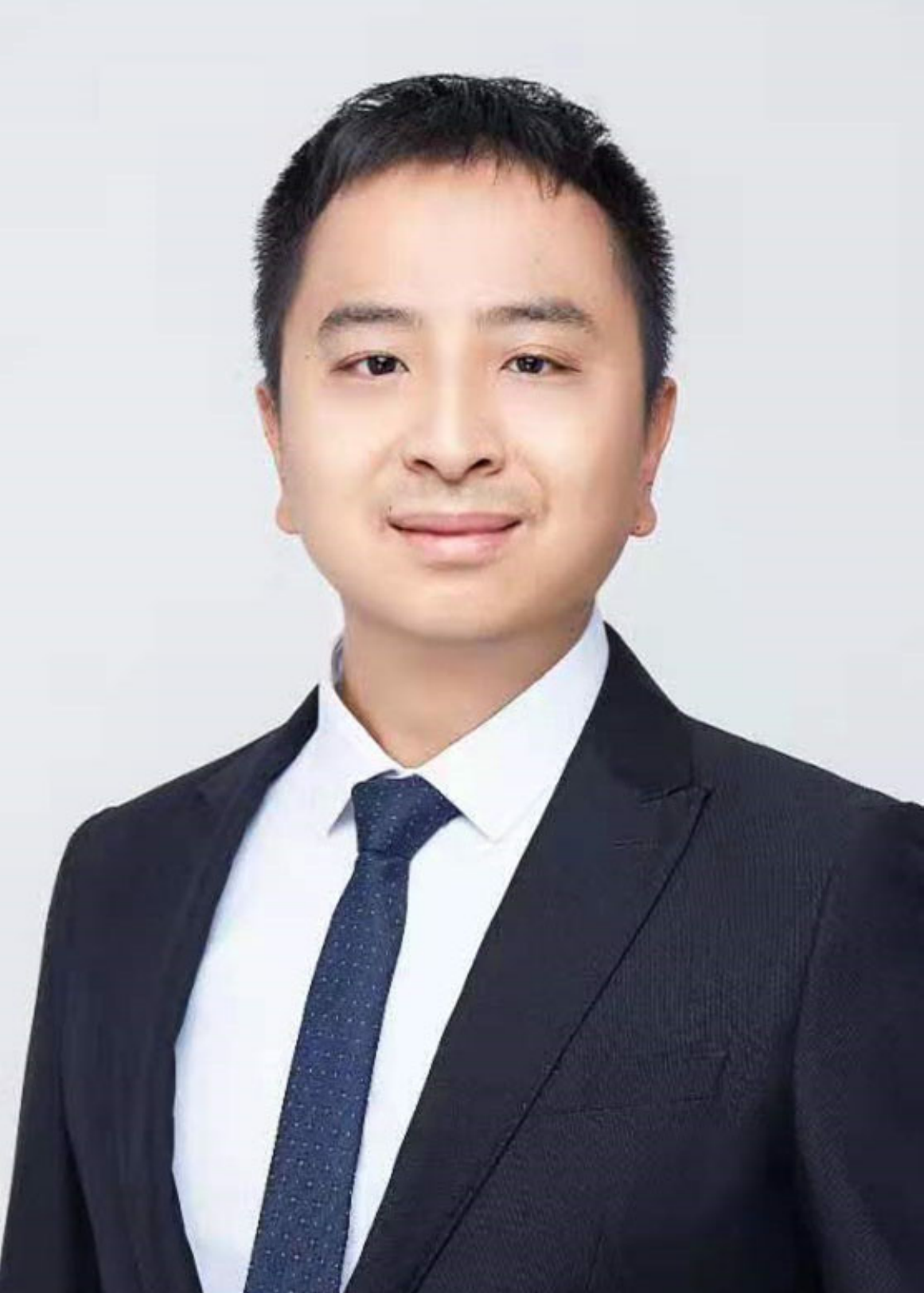}}]{Chuan Lin}
	[S'17, M'20] is currently an associate professor with the Software College, Northeastern University, Shenyang, China.
	He received the B.S. degree in Computer Science and Technology from Liaoning University, Shenyang, China in 2011, the M.S. degree in Computer Science and Technology from Northeastern University, Shenyang, China in 2013, and the Ph.D. degree in computer architecture in 2018.
	From Nove. 2018 to  Nove. 2020, he is a Postdoctoral Researcher with the School of Software, Dalian University of Technology, Dalian, China.
	His research interests include UWSNs, industrial IoT, software-defined networking.
\end{IEEEbiography}
\vspace{-5ex}
\begin{IEEEbiography}[{\includegraphics[width=1in,height=1.25in,clip,keepaspectratio]{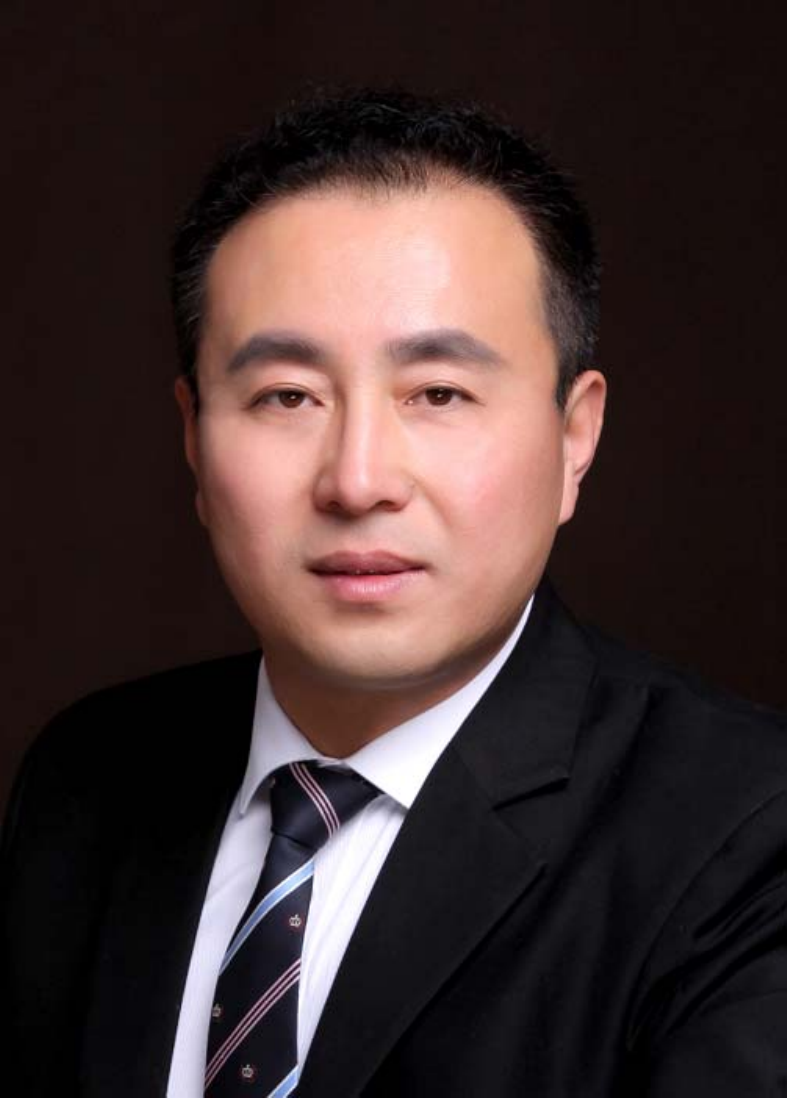}}]{Guangjie Han} [S’03-M’05-SM’18-F’22] is currently a Professor with the Department of Internet of Things Engineering, Hohai University, Changzhou, China. He received his Ph.D. degree from Northeastern University, Shenyang, China, in 2004. In February 2008, he finished his work as a Postdoctoral Researcher with the Department of Computer Science, Chonnam National University, Gwangju, Korea. From October 2010 to October 2011, he was a Visiting Research Scholar with Osaka University, Suita, Japan. From January 2017 to February 2017, he was a Visiting Professor with City University of Hong Kong, China. From July 2017 to July 2020, he was a Distinguished Professor with Dalian University of Technology, China. His current research interests include Internet of Things, Industrial Internet, Machine Learning and Artificial Intelligence, Mobile Computing, Security and Privacy. Dr. Han has over 500 peer-reviewed journal and conference papers, in addition to 160 granted and pending patents. Currently, his H-index is 81 and i10-index is 387 in Google Citation (Google Scholar). The total citation count of his papers raises above 23400 times. 
Dr. Han is a Fellow of the UK Institution of Engineering and Technology (FIET). He has served on the Editorial Boards of up to 10 international journals, including the IEEE TII, IEEE TCCN, IEEE TVT, IEEE TNSM, IEEE Systems, etc. He has guest-edited several special issues in IEEE Journals and Magazines, including the IEEE JSAC, IEEE Communications, IEEE Wireless Communications, Computer Networks, etc. Dr. Han has also served as chair of organizing and technical committees in many international conferences. He has been awarded 2020 IEEE Systems Journal Annual Best Paper Award and the 2017-2019 IEEE ACCESS Outstanding Associate Editor Award. He is a Fellow of IEEE.
\end{IEEEbiography}

\vspace{-5ex}
\begin{IEEEbiography}[{\includegraphics[width=1in,height=1.25in,clip,keepaspectratio]{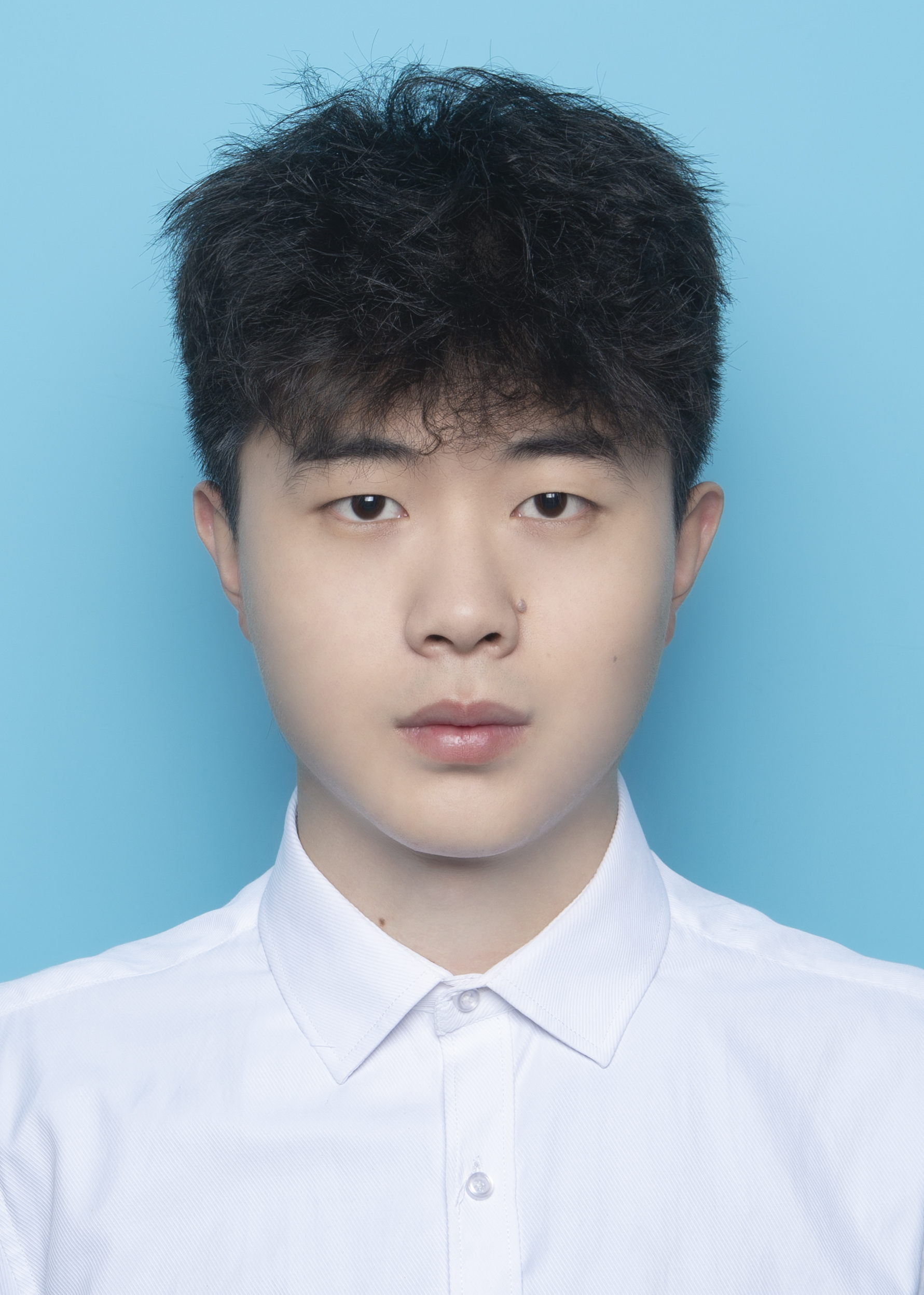}}]{Shengchao Zhu}
(Student member, IEEE) received his B.S. degree in Internet of Things Engineering
from Hohai University, Changzhou, China, in 2023. He is currently pursuing the Ph.D. degree with the Department of Computer Science and Technology at Hohai University, Nanjing, China. His current research interests include swarm intelligence, swarm ocean, Multi-Agent Reinforcement Learning.
\end{IEEEbiography}

\vspace{-5ex}
\begin{IEEEbiography}[{\includegraphics[width=1in,height=1.25in,clip,keepaspectratio]{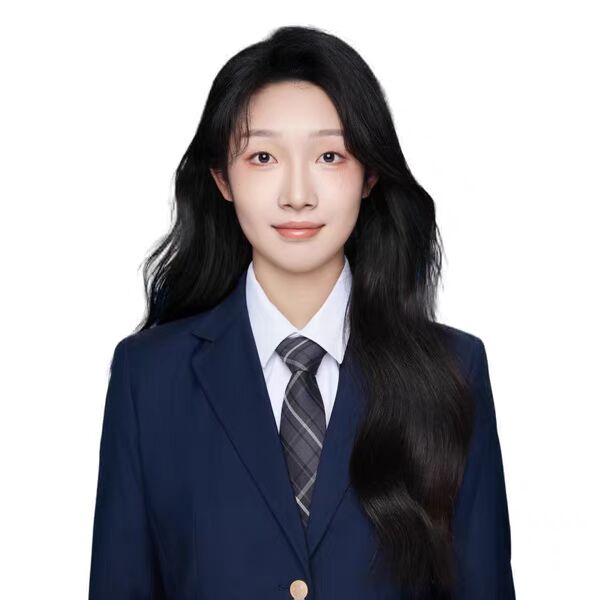}}]{Ruoyuan Wu}
is currently pursuing a Bachelor's degree at the Software College, Northeastern University, Shenyang, China. Her current research interests include computer network and machine learning.
\end{IEEEbiography}
\vspace{-5ex}

\vspace{-5ex}
\begin{IEEEbiography}[{\includegraphics[width=1in,height=1.25in,clip,keepaspectratio]{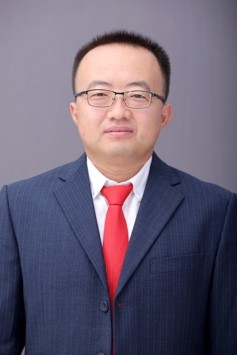}}]{Tongwei Zhang}
is currently a Professor with the Institute of Marine Science and Technology, Shandong University, Qingdao, China. He received B.S. and Ph.D. degrees in underwater acoustic engineering from Northwestern Polytechnical University, Xi'an, China, in 2006 and 2011, respectively. From 2017 to 2018, he was a Visiting Scholar with the School of Electrical and Computer Engineering, Georgia Institute of Technology, Atlanta, GA, USA. His research interests include underwater acoustic positioning, underwater acoustic communication, cooperative navigation, underwater PNT, and underwater sensor networks.

\end{IEEEbiography}
\vspace{-5ex}

\vspace{-5ex}
\begin{IEEEbiography}[{\includegraphics[width=1in,height=1.25in,clip,keepaspectratio]{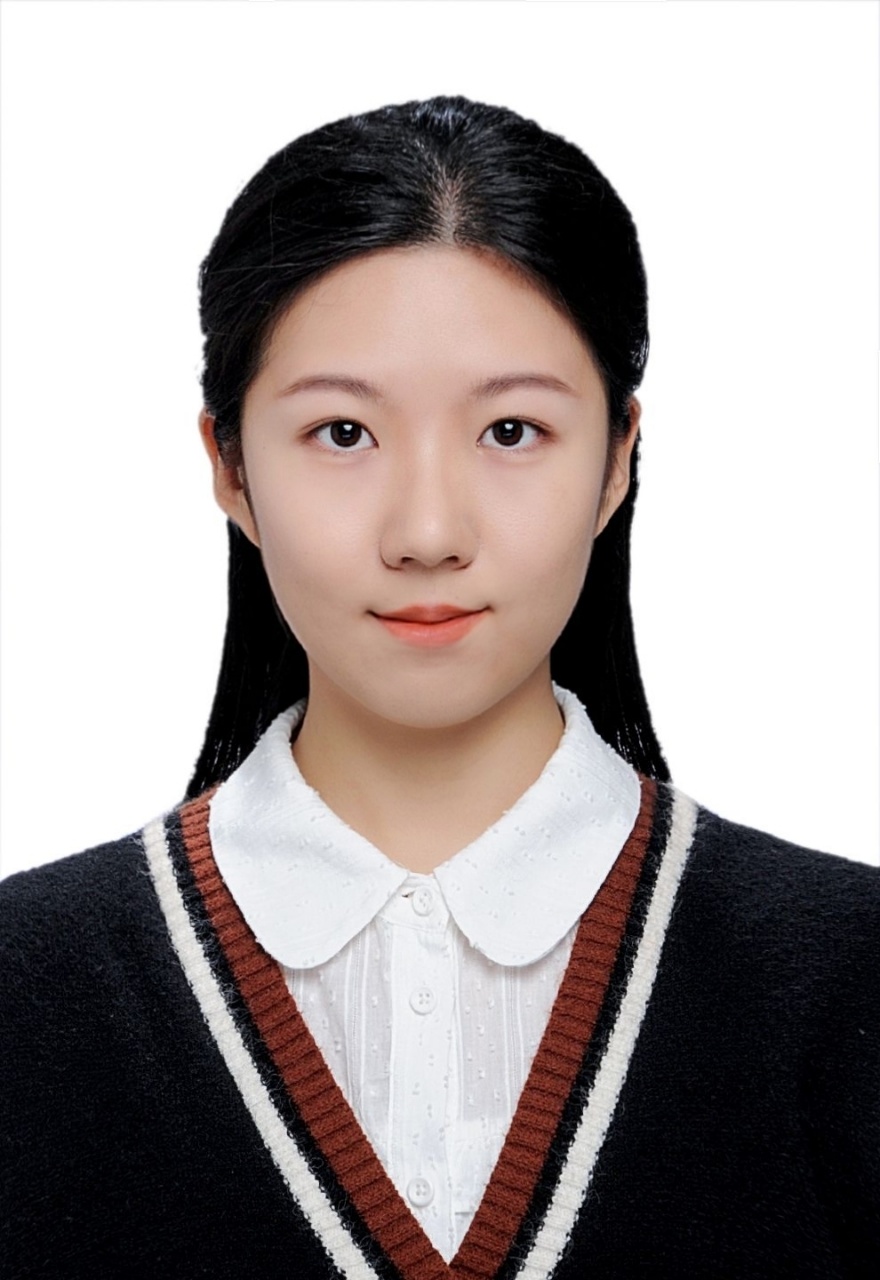}}]{Jialu Tian}
is pursuing a Bachelor's degree in School of Software, Northeastern University, Shenyang, China. Her current research interests include reinforcement learning, computer vision, and geometric modeling.
\end{IEEEbiography}

\end{document}